\documentclass[fleqn,usenatbib]{mnras}

\usepackage{newtxtext,newtxmath}

\usepackage[T1]{fontenc}

\DeclareRobustCommand{\VAN}[3]{#2}
\let\VANthebibliography\thebibliography
\def\thebibliography{\DeclareRobustCommand{\VAN}[3]{##3}\VANthebibliography}

\usepackage{subfig}

\usepackage{graphicx}	
\usepackage{amsmath}	
\usepackage{newtxtext,newtxmath}	
\usepackage{multirow}
\usepackage{tabularx}
\usepackage{arydshln}

\title[]{Localised ejection of dust and chunks on comet 67P/Churyumov-Gerasimenko: testing how comets work}

\author[N. Attree et al.]{
N. Attree$^{1}$$^{,}$$^{2}$\thanks{E-mail: attree@iaa.es},
C. Schuckart$^{2}$,
D. Bischoff$^{2}$
B. Gundlach$^{3}$,
J. Blum$^{2}$
\\
$^{1}$Instituto de Astrofísica de Andalucía (CSIC), Glorieta de la
Astronomía s/n. 18008 Granada, Spain\\
$^{2}$Institut f\"ur Geophysik und Extraterrestrische Physik, Technische Universit\"at Braunschweig, Mendelssohnstr. 3, 38106 Braunschweig, Germany\\
$^{3}$Institut f\"ur Planetologie, Universit\"at M\"unster, Wilhelm-Klemm-Str. 10, 48149, M\"unster, Germany
}

\date{Accepted XXX. Received YYY; in original form ZZZ}

\pubyear{2024}

\begin{document}
\label{firstpage}
\pagerange{\pageref{firstpage}--\pageref{lastpage}}
\maketitle

\begin{abstract}
We extend an existing thermophysical activity model of comet 67P/Churyumov-Gerasimenko to include pressure buildup inside the pebbles making up the nucleus. We test various quantities of H$_{2}$O and CO$_{2}$, in order to simulate the material inside and outside of proposed water enriched bodies (WEBs). We find that WEBs can reproduce the peak water flux observed by Rosetta, but that the addition of a time-resolved heat-flow reduces the water fluxes away from perihelion as compared to the previously assumed equilibrium model. Our modelled WEBs eject dust continuously but with a rate that is much higher than the observed erosion and mass-loss, thus requiring an active area smaller than the total comet surface area or very large quantities of dust fallback. When simulating the CO$_{2}$-rich non-WEB material, we only find the ejection of large chunks under specific conditions (e.g.~low diffusivities between the pebbles or intense insolation at southern summer), whilst we also find CO$_{2}$ outgassing rates that are much greater than observed. This is a general problem in models where CO$_{2}$ drives erosion, alongside difficulties in simultaneously ejecting chunks from deep whilst eroding the surface layer. We therefore conclude that ejection of chunks by CO$_{2}$ must be a localised phenomenon, occurring separately in space or time from surface erosion and water emission. Simulating the global production rates of gas, dust, and chunks from a comet thus remains challenging, while the activity mechanism is shown to be very sensitive to the material structure (i.e.~porosity and diffusivity) at various scales.
\end{abstract}



\section{Introduction}
\label{sec:introduction}

Comets contain some of the most primitive Solar System material accessible to investigation. Cometary activity, driven by the sublimation of volatiles near the surface, processes, removes, and reveals this material to varying degrees that requires thermophysical modelling to understand. In particular, the so-called cohesion bottleneck or activity paradox, where the outgassing pressure struggles to overcome the cohesion or tensile strength holding the surface grains together, is still not fully understood \citep{Keller1993, Blum2014, Jewitt2019,bischoff2023}. Likewise, the ejection of large, decimetres- to metres-sized chunks, as seen at several comets \citep{A'Hearn2011, Lemos2023, Lemos2024, Pfeifer2024, Shi2024}, appears to require high gas pressures at relatively large depths. Meanwhile, the sublimation of water, the main driver of cometary activity close to the Sun, should be concentrated near the surface. Various models have been proposed to address the above problems. The cohesion can be reduced by increasing the size of the relevant grains \citep{Skorov.2012}, which has led to the proposal that comets are composed of mm-cm-sized 'pebbles' \citep{Blum2017}, which form in the protoplanetary disk. Meanwhile, ice species with higher volatility than water, such as CO$_{2}$, are often invoked to explain the ejection of large chunks.

Attempts to model these processes numerically have proved difficult, however. For comet 67P/Churyumov-Gerasimenko (hereafter referred to as 67P), \citet{Gundlach.2020} presented a detailed thermophysical model, taking into account varying thermal conductivity due to contacts within and between pebbles as well as radiative transport, and calculating pressure build-up due to H$_{2}$O and CO$_{2}$ sublimation with an analytical solution for gas diffusion. This was applied only to the constant illumination conditions at the comet's southern hemisphere at perihelion, but was extended to varying illumination conditions in \citet{bischoff2023}. The effects of varying the conductivity, diffusivity, and areal emission rate were studied in this paper, while various dust ejection scenarios were modelled. These included criteria based on the build-up of gas pressure as well as the draining of volatiles from layers. However, it was found that in none of these scenarios could the emission rates of water, CO$_{2}$, and dust simultaneously be matched with the time-varying measurements of 67P made by Rosetta. In cases with large diffusivity, pressure build-up was insufficient to overcome even very low tensile strengths, while at lower diffusivity, pressure was increased, but the H$_{2}$O emission rate fell below the measurements. Additionally, in all cases where CO$_{2}$ controlled the ejections (either in draining or pressure build-up), it also dominated the outgassing rate, exceeding that of water, unlike in the Rosetta measurements. This was in contrast to the results for constant illumination conditions in \citet{Gundlach.2020}.

Meanwhile, \citet{Davidsson2022} applied another detailed thermal model, described in \citet{Davidsson2021MNRAS}, to 67P. This model includes a full numerical treatment of  gas flow as well as many other effects, but does not attempt to explain the ejection of material as relating to the pressure build-up, instead prescribing a surface erosion rate. \citet{Davidsson2022} found that several parameters, such as those controlling diffusivity had to vary before and after 67P's perihelion in order to fit the outgassing data, and that reasonable fits could be found with this variation plus an input erosion rate corresponding to equal masses of water and dust emission.

In order to address some of the problems described above, a further model of cometary material has been developed by \citet{Fulle2019}, refined in subsequent papers \citep{Fulle2020, Ciarniello2022, Ciarniello2023}, and applied in order to explain outgassing rates, isotopic ratios, and certain erosion rate measurements on 67P and other comets \citep{Fulle2020b, Fulle2021, Fulle2022}. This model, that we will refer to as the \citet{Fulle2019} model, proposes that the $\sim$cm-sized pebbles composing comets themselves contain an hierarchical structure of clusters of individual grains, mixed with various ices. Sublimation of the ices then happens within the grains, where it is assumed that gas diffusivity is controlled by the smallest grain-sizes and is therefore much lower than in-between pebbles. Very high pressures (hundreds or even thousands of Pa) can then build up. Tensile strengths within the pebbles, meanwhile, are still relatively weak thanks to few connection points due to the high porosity. In a recent paper, \citet{Kreuzig.2024} showed that outgassed dust-ice pebbles can possess internal tensile strengths below 100~Pa. In this way the cohesion bottleneck can be overcome and small, sub-pebble, dust particles can be broken off and ejected by the gas flow. 

The model of \citet{Fulle2019} has been used to explain the total outgassing rate of 67P \citep{Ciarniello2023}, and it has also been extended in \citet{Ciarniello2022} to also describe surface colour changes, by proposing two different types of the above hierarchical pebble. Water Enriched Bodies (WEBs) are approximately metre-sized concentrations of water-rich pebbles (with dust-to-water-ice mass ratios of $\delta\sim2$; \citealp{ORourke2020}) which are exposed on the surface as blue patches \citep{Fornasier2023} by CO$_{2}$ driven activity in the rest of the, water-poor (dust-to-water-ice mass ratios of $\delta\sim50$; \citealp{Fulle2021}) but CO$_{2}$-rich, non-WEB material. The seasonal blueing of the surface towards perihelion is then explained by an increase in CO$_{2}$ activity, modelled empirically in \citet{Ciarniello2023}, exposing more WEBs \citep{Ciarniello2022}. Water outgassing, meanwhile, follows the  \citet{Fulle2019} model rates as surface material is either progressively eroded in WEBs, or refreshed by ejection of large (10s of cm to $\sim$metres-sized) chunks by CO$_{2}$ in non-WEBs.

The \citet{Fulle2019} and \citet{Ciarniello2022} models thus present an attractive solution to the cohesion bottleneck, as well as a general explanation of 67P's outgassing by a relatively simple model. However, it is this simplicity that needs to be tested. The gas emission rates computed in \citet{Fulle2020} assume a steady-state thermal equilibrium for the surface material, while in reality diurnal insolation changes will ensure this is not always the case. Similarly, \citet{Ciarniello2023} only model CO$_{2}$ in an empirical way, while it is vital to understand its contribution to outgassing, ejections, and the exposure of fresh WEBs in the \citet{Ciarniello2022} model. Thus, a more detailed, numerical treatment is required.

Therefore, we have extended the thermophysical model described in the previous work \citep{bischoff2023} to include a description of sub-pebble structure and pressure build-up, in order to model the emission of H$_{2}$O, CO$_{2}$, dust, and chunks in a fully time-dependent way. The numerical implementation is described below in Section \ref{sec:TPM}. The results of various simulation runs are presented in Section \ref{sec:results}, discussed in Section \ref{sec:discussion}, and conclusions are drawn in \ref{sec:conclusions}.

\section{Thermophysical model}
\label{sec:TPM}

\begin{table*}
\caption{Definition of the variables used here: top values are fixed constants, while those below the line are variable within the model. See \citet{bischoff2023} and \citet{Gundlach.2020} for values of other physical parameters.}
\begin{tabular}{llccc}
\hline
Symbol & Parameter & Value & Unit & Reference \\ \hline
$\alpha$ & Outgassing area per depth & 1 & m$^{2}$ & Sect.~\ref{sec:TPM:Basis} \\
$a_{m}$ & Minimum pore diameter inside pebbles & 20 & nm & \citet{Fulle2020} \\
$d_{p}$ & Minimum grain diameter inside pebbles & 100 & nm & \citet{Fulle2020} \\
$k$ & Boltzmann constant & $1.38 \times 10^{-23} $&$ \mathrm{J \, K^{-1}}$ & - \\
$\rho_{bulk}$ & Bulk density & $532 $&$ \mathrm{kg \, m^{-3}}$ & \citet{Jorda.2016} \\
R & Pebble radius & 5 & mm & \citet{Fulle2020} \\
\hline
b & Diffusion parameter & & m & Sect.~\ref{sec:method:diffusivity} \\
c & Specific heat capacity at specified point & & J kg$^{-1}$ K$^{-1}$ & \citet{Gundlach.2020} \\
d & Pore diameter in specified model & & m &  Sect.~\ref{sec:method:diffusivity} \\
$\delta$ & Dust-to-total-ice mass ratio & &  & - \\
$\delta_{i}$ & Dust-to-ice-species $i$ mass ratio & &  & - \\
$\epsilon$ & Porosity at specified point & &  & - \\
$\eta$ & Diffusion efficiency function & & & Eqn.~\ref{eq:eta} \\
$j_{leave}$ & Total outwards outgassing flux & & kg s$^{-1}$ m$^{-2}$ & Sect.~\ref{sec:TPM:Basis} \\
$j_{inwards}$ & Total inwards outgassing flux & &  kg s$^{-1}$ m$^{-2}$ & Sect.~\ref{sec:TPM:Basis} \\
$\lambda$ & Thermal conductivity at specified point & & W m$^{-1}$ K$^{-1}$ &  Eqn.~\ref{eq:pebble_conductivity} \\
$P_{in}$ & Pressure inside pebbles at specified point & & Pa & Eqn.~\ref{eq:Pin} \\
$P_{out}$ & Pressure between pebbles at specified point & & Pa & Eqn.~\ref{eq:Pout} \\
q & Outgassing flux from pebbles & &  kg s$^{-1}$ m$^{-2}$ & Eqn.~\ref{eq:q} \\
$q_{out}$ & Outgassing flux between pebbles & &  kg s$^{-1}$ m$^{-2}$ & Eqn.~\ref{eq:qout} \\
$\rho$ & Density at specified point & &$ \mathrm{kg \, m^{-3}}$ & - \\
$\sigma$ & Tensile strength &  & Pa & Sect.~\ref{sec:method:strength} \\
T & Temperature at specified point & & K & Eqn.~\ref{eq:heat_transport_eq} \\
t & Time in simulation & & s & - \\
v & Gas thermal velocity at specified point & & m s$^{-1}$ & Sect.~\ref{sec:method:diffusivity} \\
z & Depth below surface &  & m & - \\
\hline
\end{tabular}
\label{Tab:1_Parameters}
\end{table*}

\subsection{Model of internal pebble pressure}
\label{sec:method:pinside}

We first briefly describe the model of \citet{Fulle2019} and \citet{Fulle2020}. It is assumed that the comet is made up of pebbles of radius $R=5$ mm with a heterogeneous substructure composed of smaller aggregates with a power-law size-distribution down to their smallest component grains of uniform particle diameter $d_{p}=100$ nm. Upon the further assumptions of zero pressure outside the pebble (which will be returned to later), and uniformly distributed ice sources inside (coating the small grains, or distributed among them as similarly sized ice-grains), equations are constructed in \citet{Fulle2019} and \citet{Fulle2020} to represent the average pressure inside, and the outgassing flux from, a single pebble of uniform temperature $T$:
\begin{equation}
    P_{in} \approx P_{sat} = A e^{-B/T}
    \label{eq:Pin}
\end{equation}
and
\begin{equation}
    q = \frac{35a_{m}P_{in}}{3R} \sqrt{\frac{2m}{\pi kT}},
    \label{eq:q}
\end{equation}
where $A,B$ and $m$ are the sublimation constants and molecular mass of the relevant ice species (H$_2$O and CO$_2$, values as used in \citet{Gundlach.2020}), $k$ is the Boltzman constant, and $a_{m}$ is the average pore radius amongst the smallest grains. For consistency with \citet{Fulle2020} we use $a_{m}=0.2d_{p}=20$ nm, but we will return to discuss diffusivity in more detail below. Table \ref{Tab:1_Parameters} below gives an overview of the parameters and constants used here.

A certain volume of bulk material composed of such sublimating pebbles will then develop a pressure between them, $P_{out}$ in Pa, and an outgassing flux, $q_{out}$ in kg s$^{-1}$ m$^{-2}$, which is damped by the gas diffusion between (rather than inside) the pebbles. This pressure is much smaller than the internal pressure (hence its exclusion from the equations above), due to the much higher diffusion facilitated by the larger pore spaces, which are proportional to the pebble size $d=2R$. This damping was expressed in the previous work \citep{bischoff2023} using the efficiency factor \citep{Gundlach.2020}
\begin{equation}
    \eta = \left( 1 + \frac{z}{b} \right)^{-1} ,
    \label{eq:eta}
\end{equation}
for a certain depth, $z$, below the surface, and the $b$ parameter describing the material's gas diffusivity. The $b$ parameter is equal to the number of particle layers required to reduce the outgassing rate by a factor of 2. Using this factor, the pressure in-between pebbles and the total outwards gas-flux of the material are written
\begin{equation}
    P_{out} = q \sqrt{\frac{2\pi kT}{m}} (1 - \eta),
    \label{eq:Pout}
\end{equation}
 as in \citet{Fulle2020} Equation 16, and
\begin{equation}
    q_{out} = P_{out} \sqrt{\frac{m}{2 \pi k T}} \, \eta \alpha,
    \label{eq:qout}
\end{equation}
respectively, where $\alpha$ is the total outgassing area in the volume being considered. In \citet{Fulle2020}, only outgassing through the top surface of the upper pebbles is considered, whereas as described below (and as in \citet{bischoff2023} and \citet{Gundlach.2020}) here we consider inwards gas flow as well, so that our $q_{out}$is double that of \citet{Fulle2020}.

We next describe how this model is integrated into the existing one-dimensional thermophysical model developed by \citet{Gundlach.2012, Gundlach.2020} and used in \citet{Bischoff.2021, Burger.2022}; and \citet{bischoff2023}.

\subsection{Numerical implementation}
\label{sec:TPM:Basis}

The model, which is described in detail in \citet{Gundlach.2020}, and \citet{bischoff2023}, solves the heat-transfer equation for temperature $T$ at time $t$ and depth $z$
\begin{equation}
    \rho(z) c(z) \frac{dT(z,t)}{dt} = \frac{d}{dz} \left[ \lambda(T(z),z) \frac{dT(z,t)}{dz} \right] + Q(T(z),z)
    \label{eq:heat_transport_eq}
\end{equation}
using the finite difference method and the forward difference scheme with constant $dz=R$ and $dt=10$ s. The thermal conductivity $\lambda(T(z),z)$ is temperature dependent and given by the sum of the network conductivity between contacting pebbles and the radiative component:
\begin{equation}
    \lambda = \lambda_{\mathrm{net}}(R) + \lambda_{\mathrm{rad}}(R,T) = \lambda_{\mathrm{net}}(R) + \lambda_{\mathrm{rad}}(R)\, \left( \frac{T}{\mathrm{1~K}} \right)^3,
    \label{eq:pebble_conductivity}
\end{equation}
where the parameters $\lambda_{\mathrm{net}}(R)$ and $\lambda_{\mathrm{rad}}(R)$ are described by the equations in \citet{Gundlach.2020}: section A1.1. For $\sim$cm-sized pebbles $\lambda_{\mathrm{rad}}$ dominates over the network conductivity at temperatures where most activity happens.

The density $\rho(z)$ and heat capacity $c(z)$ of a numerical layer are also variable, depending on the time-varying local volume fraction of dust, H$_2$O, CO$_2$ and vacuum, so that heat-capacity is also computed at each time-step using the equation in \citet{Gundlach.2020}: section A1.1. 

$P_{in}$, $P_{out}$, $q$, and $q_{out}$ are then computed at each time and depth step using Equations \ref{eq:Pin}, \ref{eq:Pout}, \ref{eq:q}, and \ref{eq:qout}. In layers without any ice we assume that the pressure vanishes. In Eqn.~\ref{eq:qout}, $\alpha=1$ m$^{2}$ is set as approximately equal to the total outgassing area of a one square-metre numerical layer of thickness $dz = R$ filled with cm pebbles at a packing fraction of 0.6. As noted in the previous paper, this parameter was found to have little influence on the outgassing and activity for a reasonable range of values.

As in \citet{Gundlach.2020}, half of the sublimating gas in each layer is assumed to flow outwards ($j_{\mathrm{leave}}=q_{out}/2$), and is summed over all layers to find the total outgassing rate per time-step ($q_{H_{2}O}$ and $q_{CO_{2}}$). The remaining gas is assumed to flow inwards ($j_{\mathrm{inward}}=q_{out}/2$) and is then distributed over a number of deeper layers assuming that these molecules will condense on the dust and ice surfaces below due to the generally decreasing temperature with depth. As before, we neglect re-condensation above the sublimation front due to nighttime temperature inversions as a minor contribution to the energy balance and pressure build-up, and, as before, we test a number of different downwards redistribution coefficients and find the results insensitive to the specifics of this inwards diffusion. The energy source and sink term $Q$ in Equation \ref{eq:heat_transport_eq} can then be computed by summing
\begin{equation}
    Q = \Lambda \, (j_{\mathrm{inward}} - q_{out}) \, dt \,
    \label{eq:latent_heat}
\end{equation}
over each species (H$_2$O and CO$_2$) with latent heat, $\Lambda$.

The same boundary and initial conditions as used in \citet{bischoff2023} are applied here, i.e.~insolation for different latitudes on a spherical comet with 67P's area-equivalent radius and orbit (computed using the SPICE kernels), outgoing thermal radiation by the Stefan-Boltzmann law, and an initial temperature of $50\,\mathrm{K}$ throughout the body, which is also the lower boundary condition. As before, the pebble radius is fixed at $R = 5 \,\mathrm{mm}$, while the dust-to-total ice mass-ratio, $\delta$, and CO$_2$-to-H$_{2}$O ice mass-ratios are adjusted to fit the estimates for WEB and non-WEB material as described below. The bulk density is fixed at $532 \,\mathrm{kg/m^3}$, which determines the actual amounts of CO$_2$, H$_{2}$O, and refractories in each layer, and the other parameters are the same as in \citet{Gundlach.2020} and \citet{bischoff2023}. The code is implemented using the Numba high-performance Python compiler \citep{Numba} and parallelisation, leading to a significant speed increase over the previous paper.

In \citet{bischoff2023}, various criteria were used to determine the ejection of layers because the model generally only succeeded in building up very small pressures with-which it was difficult to overcome comets' presumed tensile strengths. As we will see, the model of \citet{Fulle2020} has no such problems, so here we only use the pressure ejections criteria; i.e.~ejection of layers when the summed vapour pressure between the pebbles ($P_{out}$) exceeds the tensile strength of the layer. For the first numerical layer a special ejection criterion is used to simulate the gradual erosion of the surface pebbles into small dust, which, along with the tensile strength, is discussed in detail below.

\subsection{Strength and erosion of the first layer}
\label{sec:method:strength}

A special ejection criterion is necessary for the surface pebbles because preliminary experiments showed that using the simple layer-ejection criteria with $P_{in}$ in the first layer resulted in an unrealistically high ejection rate. This is essentially a problem of numerical resolution: gas pressure should build up downwards through the pebble in a gradual way, following the heat-wave and ejecting progressively deeper constituent grains in a semi-random way (see discussion in section 2 of \citet{Fulle2021}), until it reaches the equilibrium described by the equations of \citet{Fulle2020}. With a coarse numerical resolution of half a pebble diameter, this gradual evolution cannot be simulated and, with the simple ejection criteria, entire pebbles are immediately ejected as soon as the comet becomes active, resulting in an extremely high erosion rate and the impossibility of heating the surface beyond the activation temperature ($\sim205$ K: \citet{Fulle2020}).

As a full numerical simulation of the internal pressure distribution of a non-isothermal surface pebble is beyond the scope of this thermal model, we instead model the gradual erosion by assuming, as in \citet{Fulle2020}, that it progresses at the speed of the thermal wave through the first layer 
\begin{equation}
    q_{d} = \frac{\lambda_{0}}{R c_{0}}, 
    \label{eq:heat_wave}
\end{equation}
where $q_{d}$ is the mass flux of sub-pebble dust (unresolved but with a size distribution given by the pressure/strength curves as in \citet{Fulle2020}, Figure 3), and $\lambda_{0}$ and $c_{0}$ are the time-dependent thermal conductivity and heat-capacity of the first layer. This mass flux is activated whenever $P_{in}$ exceeds the average pebble strength (see below), and saved so that when its sum reaches the total of one pebble, the layer is ejected as above, and the next layer becomes the active pebble layer.

As we will show in the results section, pressure build-up is by far the largest in the first numerical layer, but $P_{in}$ exceeding tensile strength can also develop in lower layers (especially when considering CO$_{2}$, see section \ref{sec:results:nonWEB}). By only applying the above erosion model to the top layer, we are making the assumption that pressure within pebbles can only eject dust, rather than simply breaking the bonds holding it together\footnote{The distinction between erosion and weathering in a geological context.}, when dust particles are free to escape, and not when buried beneath other pebbles.

\begin{figure}
\resizebox{\hsize}{!}{\includegraphics{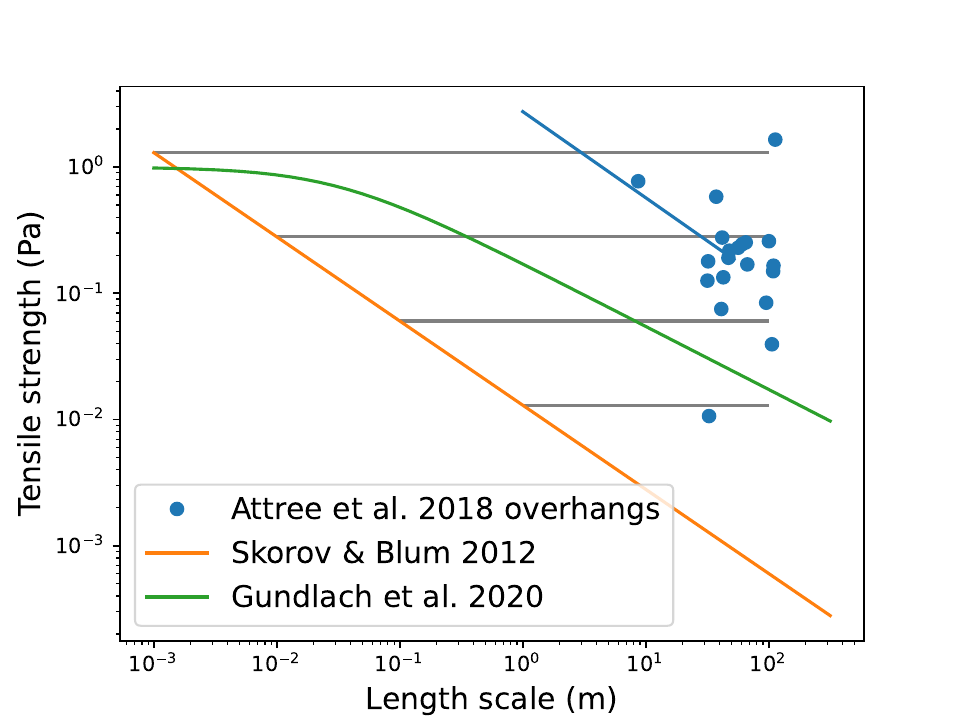}}
\caption{Tensile strength laws from \citet{Skorov.2012} and used in \citet{Gundlach.2020} for varying length-scale and for certain fixed particle sizes for \citet{Skorov.2012} (black lines). Also shown are the measurements from overhanging cliffs from \citet{Attree2018} and the scaling for ice used therein.}
\label{PLot_StrengthLengthScale}
\end{figure}

Regarding the material tensile strength, there are a number of estimates from Rosetta and other sources (see \citet{Groussin2019, Biele.2022b} for overviews), but little consensus. We choose to use the same expression as in \citet{Fulle2020} which comes from the theory of \citet{Skorov.2012}, i.e.~$\sigma = 13 z^{-2/3}$ mPa, where depth $z$ has been equated to particle size. \citet{Gundlach.2020} use a slightly different depth-dependent strength which is based on the statistical weakness model of increasing likelihood of defects with increasing scale. Equating this scale-length with depth, the two expressions are plotted together with the measurements of overhanging cliff strengths from \citet{Attree2018} in Figure \ref{PLot_StrengthLengthScale}. A real material composed of heterogeneously sized dust particles hierarchically arranged into pebbles should follow the \citet{Skorov.2012} strength law for increasing size up until the pebble size is reached, whereupon the homogeneously sized pebbles should lead to a constant strength with scale (horizontal lines shown in Fig.~\ref{PLot_StrengthLengthScale} for 1 mm, 1 cm, 10 cm, and 1 m). At even larger scales, the defects law \citep{Gundlach.2020, Biele.2022b} should take over; however, we are typically concerned with decimetre or below scales here. Due to the large uncertainties present in all the relevant parameters, and the huge range of values in the strength measurements themselves, we consider scale-varying strength laws too complex, and use the relatively simple \citet{Skorov.2012} expression for consistency with \citet{Fulle2020}. Note that this gives a low, likely minimum, estimate for the strength between pebbles, while the internal pebble strength could also be higher, and porosity dependent \citep{Kreuzig.2024}.

As shown by \citet{bischoff2023} in their Figure 1, pressures exceeding the tensile strength automatically satisfy $P>P_{grav}=\rho gz$, i.e.~the lift criteria for a block of material at depth z, for reasonable parameter choices, so that ejections do indeed lift material away.

\subsection{Diffusivity}
\label{sec:method:diffusivity}

As shown in \citet{bischoff2023}, gas diffusivity is a critical parameter in determining the sublimation rate at depth, and therefore the energy balance, the pressure build-up, and the ejections. The description of diffusivity by the efficiency function, $\eta$, and the $b$ parameter was originally made in \citet{Gundlach.2011b}, where $b = [6-7]d_{p}$ was measured experimentally for $d_{p}=2$ mm and $55\, \mathrm{\mu m}$ diameter particles at a porosity of $\epsilon=0.4$. Subsequently, a lower value of $b = d_{p}$ was used in modelling \citep{Gundlach.2020, bischoff2023}.

In \citet{Gundlach.2020} section A1.2, a relationship was derived between the efficiency function and the Knudsen diffusion coefficient, $D_{K}$. Subsequently, \citet{Macher2024} gave a full treatment of the efficiency function, or half-transmission thickness, by the use of a statistical method, deriving it as a function of particle size, as well as its relation with $D_{K}$. The two expressions in \citet{Gundlach.2020} and \citet{Macher2024} differ by a factor of porosity, $\epsilon$, which is needed, as explained in the latter publication, to account for the fact that gas flow only occurs through the pores, with a fraction equal to $\epsilon$ for a homogeneous distribution. The relation is then
\begin{equation}
    D_{K} = \frac{vb\epsilon}{4}, 
    \label{eq:knudsen_Dk}
\end{equation}
where $v=\sqrt{8kT/\pi m}=2\sqrt{2kT/\pi m}$ is the gas thermal velocity. Taking this factor of porosity into account, the relationship with the size of the assumed spherical particles is shown using the statistical argument to be
\begin{equation}
    b = \frac{d_{p}\epsilon}{2(1-\epsilon)}.
    \label{eq:knudsen_b}
\end{equation}
This model is found to be very accurate when compared to others in the literature and tested against DSMC simulations. In addition to this work, \citet{Skorov.2011, Skorov.2021, Reshetnyk.2021, Skorov.2022}, and \citet{Reshetnik.2022} also ran statistical simulations of gas transport for variously structured materials and found that the mean-free-path, MFP, of the gas molecules between collisions with the grains followed the relationships $\mathrm{MFP}\approx\epsilon/(1-\epsilon)$ in units of $d_{p}$. Using \citet{Skorov.2011}'s MFP as an estimate of b therefore results in a value that is two times larger than Eqn.~\ref{eq:knudsen_b}.

Previous works by \citet{Derjaguin} and \citet{Asaeda}, also derived theoretical models from statistical and momentum-transfer considerations, while \citet{Guettler23}, again confirmed their accuracy with DSMC simulations and laboratory experiments. \citet{Guettler23} show that the \citet{Derjaguin} result can be expressed as
\begin{equation}
    D_{K} = \frac{3\epsilon v \mathrm{MFP}}{13} = \frac{2vd_{p}}{13} \frac{\epsilon^{2}}{1 - \epsilon} ,
    \label{eq:knudsen_MFP}
\end{equation}
where 
\begin{equation}
    \mathrm{MFP} = \frac{2d_{p}}{3}\frac{\epsilon}{1-\epsilon},
    \label{eq:MFP}
\end{equation}
which is equal to two thirds of the \citet{Skorov.2011} value above, and is also equivalent to the \citet{Asaeda} expression for certain values of their parameters. Indeed, \citet{Macher2024} note that the \citet{Asaeda} expression is a simplification of the \citet{Derjaguin} result, which itself requires full knowledge of the distribution of path-lengths travelled by the gas particles, as available in simulations. Equating Eqn~\ref{eq:knudsen_MFP} with Eqn.~\ref{eq:knudsen_Dk} gives
\begin{equation}
b= \frac{8d_{p}\epsilon}{13(1 - \epsilon)},
    \label{eq:b_Asaeda}
\end{equation}
a factor of 13/16 larger than Equation \ref{eq:knudsen_b}.

In contrast with the above statistical approaches, which use spherical particles, the assumption of parallel cylindrical pores is often used in the comet literature as it allows an easier analytical approach. The Knudsen diffusion coefficient of a single cylindrical tube with pore diameter $d$ is written $D_{K}^{1} = \frac{vd}{3}$ (e.g.~\citealp{Kast2000}). Generalising this to a bulk material then typically assumes it to be composed of parallel tubes with a certain number density or bulk porosity, and a bent structure that can be described by a tortuosity, $\tau$. The total Knudsen diffusion is then written (e.g.~as in \citealp{Schweighart.2021})
\begin{equation}
    D_{K} = \frac{\epsilon}{\tau^{2}} D_{K}^{1} = \frac{vd\epsilon}{3\tau^{2}}. 
    \label{eq:knudsen_bent}
\end{equation}
The relationship between the pore diameter and the physical particle size, as well as the exact value of the tortuosity, are then complicated and porosity dependent. Typically $\tau=1-3$ and $d=d_{p}$, i.e.~the pore sizes are assumed equal to the grain-size, or, as found experimentally by \citet{Schweighart.2021} $d\sim0.5d_{p}$. In \citet{Fulle2019}, it is assumed that the average pore radius (not diameter) is 20\% of the particle diameter, i.e.~$d=0.4d_{p}$, and the Knudsen diffusion coefficient is written ignoring its porosity dependence as
\begin{equation}
    D_{K} = \frac{2d_{p}v}{15}.
    \label{eq:knudsen_Fulle}
\end{equation}
Equating this with Equation \ref{eq:knudsen_b} to obtain the appropriate b value to directly compare to the \citet{Fulle2020} model results in the porosity dependent expression
\begin{equation}
    b = \frac{8d_{p}}{15\epsilon},
    \label{eq:b_Fulle}
\end{equation}
which evaluates as $b=[0.9-1.52] d_{p}$ for $\epsilon=[0.6-0.35]$.

\begin{figure}
\resizebox{\hsize}{!}{\includegraphics{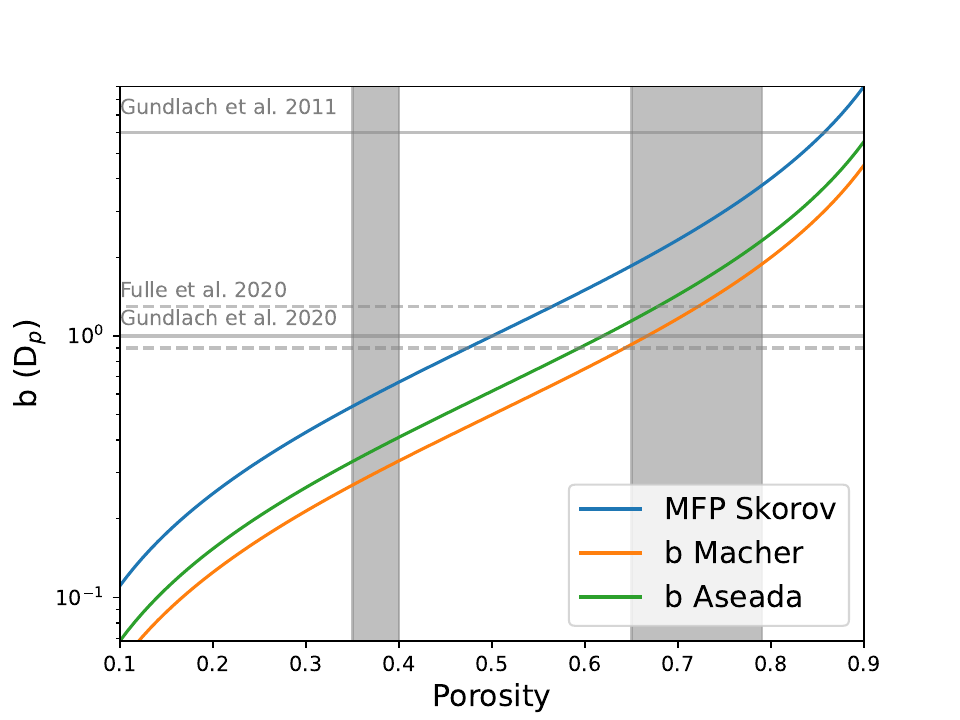}}
\caption{Mean-free-path (MFP) and the $b$ parameter in units of particle diameter and as a functions of porosity from \citet{Skorov.2022} and \citet{Macher2024}, and based on \citet{Asaeda} and \citet{Derjaguin}. Porosities of random loose- to close-packed spheres and bulk 67P \citep{Patzold} are also shown, as well as several other labelled estimates of $b$.}
\label{Plot_Tempmodels_diffusivity}
\end{figure}

Comparing the numerical values derived from experiments and theory in Figure \ref{Plot_Tempmodels_diffusivity}, it can be seen that there is some scatter between various diffusivity models, and a rather large dependence on porosity. The two grey boxes highlight the bulk porosity of 67P ($\epsilon\approx0.65-0.8$; \citealp{Patzold}) and the approximate range of randomly packed spheres. \citet{Fulle2020} suggests random close packing ($\epsilon=0.35$) at each of three constituent levels of hierarchy to arrive at a total bulk porosity towards the top end of the measurements. The total porosity within the pebbles (two layers of hierarchy) is then $1-(1-0.35)^{2}\approx0.6$. \citet{Gundlach.2020} and \citet{bischoff2023} assumed only two levels, with $\epsilon=0.6$ inside the pebbles and $\epsilon=0.4$ outside. Gas flow at each of these scales should interact with the relevant constituent particles of size $d_{p}$ according to the porosity at that level, so it can be seen that the exact structure (i.e~number of hierarchical levels and the particle sizes and porosities at each) is very important in determining the gas flow.

For gas flow within the pebbles, it appears that $b\approx d_{p}$ is a reasonable approximation for the diffusivity, so that the results of the \citet{Fulle2020} model can be used for internal pebble pressures. Diffusion between pebbles, meanwhile, may be slightly lower than previously assumed, with b values of a few tenths of a pebble diameter being appropriate. For ease of comparison with the previous results, we will begin our numerical simulations assuming $b=d_{p}$, and discuss the effects of varying the diffusivity later.

\section{Results}
\label{sec:results}

We ran a number of simulations with the parameters fixed as in Table \ref{Tab:1_Parameters}, starting with $b=d_{p}$ and varying the dust-to-ice mass ratios of the two ice components to approximate the WEB and non-WEB material. Each simulation is initialised with the comet at aphelion with uniform 50 K temperature throughout, and is run for three orbital cycles to ensure repeated activity. Unless otherwise noted, we display the third of the, near-identical, repeats.

\subsection{WEB material}
\label{sec:results:WEB}

Inside the Water Enriched Bodies, $\delta\sim2$ \citep{ORourke2020} and there should be negligible CO$_{2}$. We therefore set $\delta=\delta_{H_{2}O}=2$ and $\delta_{CO_{2}}=0$ and use the internal pebble pressure and ejection criteria described in Sections \ref{sec:method:pinside} and \ref{sec:method:strength} and $b=d_{p}$ for diffusivity outside the pebbles.

\begin{figure}
\subfloat{\includegraphics[scale=0.5]{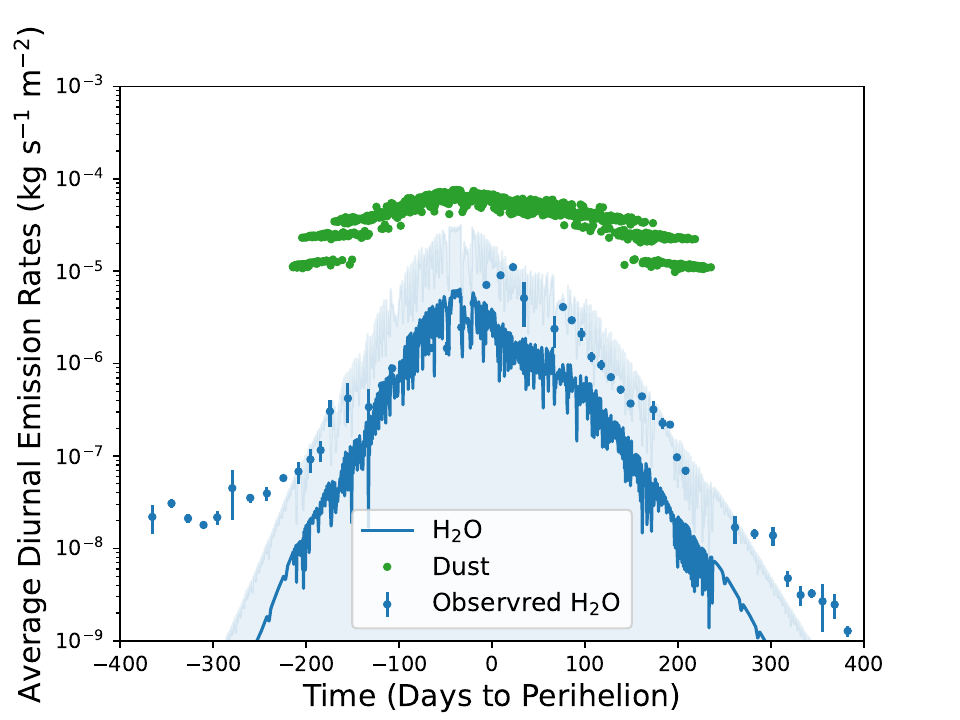}}%
\qquad
\subfloat{\includegraphics[scale=0.5]{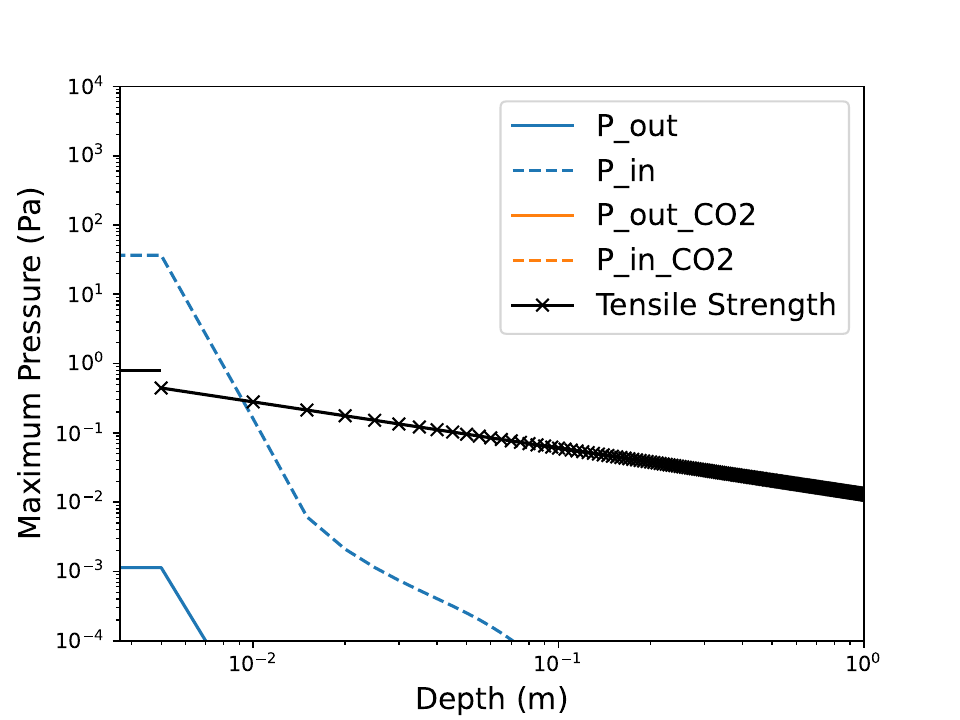}}%
\qquad
\caption{Simulation results for the \citet{Fulle2019} model WEB material. Top: average diurnal gas and dust emission rates, and their variation, during the Rosetta period as compared to the scaled observations. Bottom: Maximum outgassing pressure from different ice species in the subsurface, inside and in-between pebbles.}
\label{Plot:WEB}
\end{figure}

Figure \ref{Plot:WEB} shows the resulting diurnally averaged gas and dust fluxes with time during the Rosetta period for a one square-metre patch located on the equator of a spherical representation of 67P. Also shown are the observations \citep{Laeuter2020}, scaled down by 67P's total surface area of $5\times10^{7}$ m$^{2}$ as in \citet{bischoff2023}. As before, we do not seek to exactly match the production curves, which would be impossible for a single patch, but to ensure that the instantaneous variations around the daily rate (shaded in light blue) encompasses the measurements for this representative patch. For this patch at zero longitude and latitude, the maximum insolation is slightly before perihelion as can be seen in the peak outgassing.

Overall, the WEB material behaves as described by the \citet{Fulle2020b} model, producing roughly the correct magnitude of water emission as well as continuous dust activity within $\approx\pm200$ days from perihelion that repeats every orbit. Dust emission rates at perihelion are estimated by various instruments at between one and several hundred times the water outgassing rate \citep{Fulle2016, Ott2017, Moreno2017, Marschall2020}, and are therefore roughly matched here. Total erosion over one orbit is about 8 m, on the upper end of most estimates \citep{Keller2015}, but not unreasonably so. The lower panel of Fig.~\ref{Plot:WEB} shows the maximum pressure reached at each point in the subsurface, demonstrating that pressure build-up between the pebbles is indeed very low; even lower than in the previous models that did not take into account internal pebble structure, due to the reduced outgassing rate per pebble described in Eqn.~\ref{eq:Pout}. Pressure inside the surface pebble, on the other hand, reaches several tens of Pa, easily exceeding the tensile strength. The 8 m of erosion in this model therefore comes completely in the form of small, sub-pebble-sized dust-particles ejected by the first active pebble layer, i.e.~the top cm, where water is always present.

One difference with the computed production curve of \citet{Ciarniello2023} is that the thermal inertia (which is temperature dependent, but similar to the average 67P value of 50 thermal inertia units; \citealp{Gulkis2015}) leads to a delay in reaching the instantaneous emission rates of \citet{Fulle2020} that assume thermodynamic equilibrium. In the time-dependent model, more energy is needed to slowly heat up the surface material and diffuse the heat-wave into the subsurface, which reduces the amount of energy available for sublimation. The production rate away from perihelion is therefore lower than the time-independent model, and it is a challenge to reproduce the observations here.

\subsection{Non-WEB material}
\label{sec:results:nonWEB}

We now consider the addition of CO$_{2}$ by simulating the non-WEB material, which is water-ice depleted but super-volatile enriched, including plentiful CO$_{2}$. Exactly how plentiful is an open question: \citet{Fulle2022} estimate an upper limit of $\delta_{CO_{2}}=10$ in order for CO$_{2}$ to remain near the surface and dominate activity in incoming comets at heliocentric distances above 3.8 AU. However, in order for H$_{2}$O to take-over the activity here, CO$_{2}$ needs to have drained away from the surface already, so we here consider $\delta_{CO_{2}}=10$ to be a \textit{lower} limit instead. We therefore test two cases: an ice-rich $\delta\approx\delta_{CO_{2}}=2$ case, and an ice-poor $\delta\approx\delta_{CO_{2}}=10$ case. Water-content is set to $\delta_{H_{2}O}=50$ in both \citep{Fulle2021}.

In both cases, some initial activity is observed in the first pebble for CO$_{2}$ and then H$_{2}$O, but both swiftly drain away at a rate too high for pressure (inside or between pebbles) to rise high enough to restart activity. By the second orbit there is no longer any activity, and outgassing rates decline with each subsequent cycle as the volatiles drain deeper and deeper, reaching over a metre depth for CO$_{2}$ by the third orbit. This confirms the argument of \citet{Fulle2020}, that water dehydration rates exceed water-driven erosion rates for $\delta_{H_{2}O}=50$ near perihelion, while also showing that the same is true for CO$_{2}$, even with $\delta_{CO_{2}}=2$, due to its higher volatility. Unfortunately, this draining rate is even too high for deep layer chunk ejection to occur with these parameters.

It appears very difficult to produce good results when CO$_{2}$ mass is equal to or greater than that of H$_{2}$O. Considering this, we also test a case where CO$_{2}$ is a fraction of the water-ice mass, which is more analogous to the 'classical' description of cometary material used in the previous paper; \citet{Gundlach.2020, Davidsson2022}; etc.

\begin{figure}
\subfloat{\includegraphics[scale=0.5]{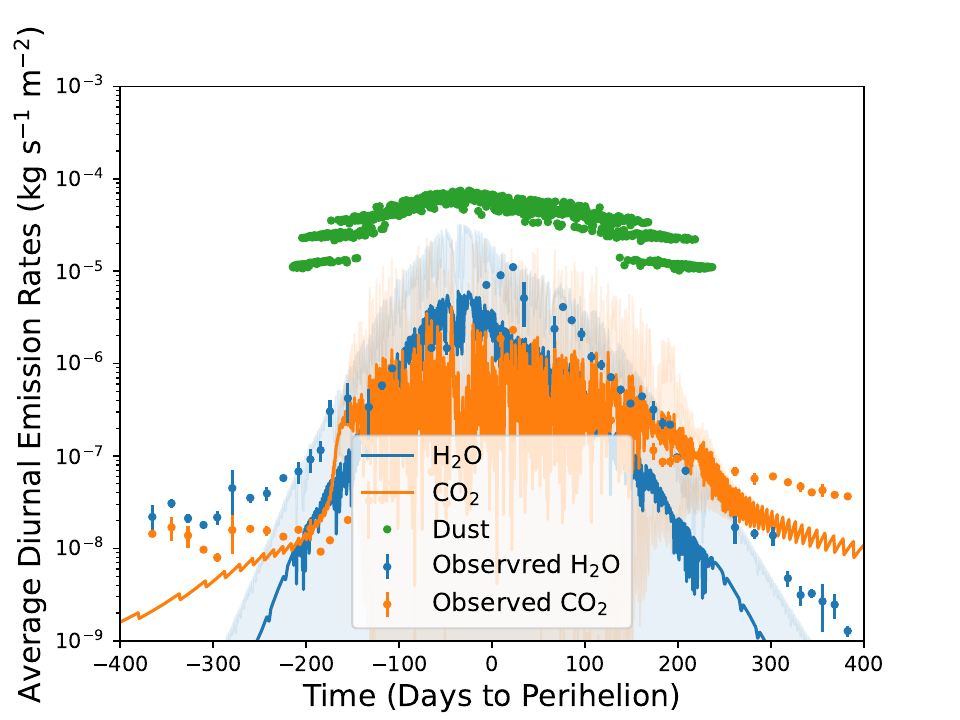}}%
\qquad
\subfloat{\includegraphics[scale=0.5]{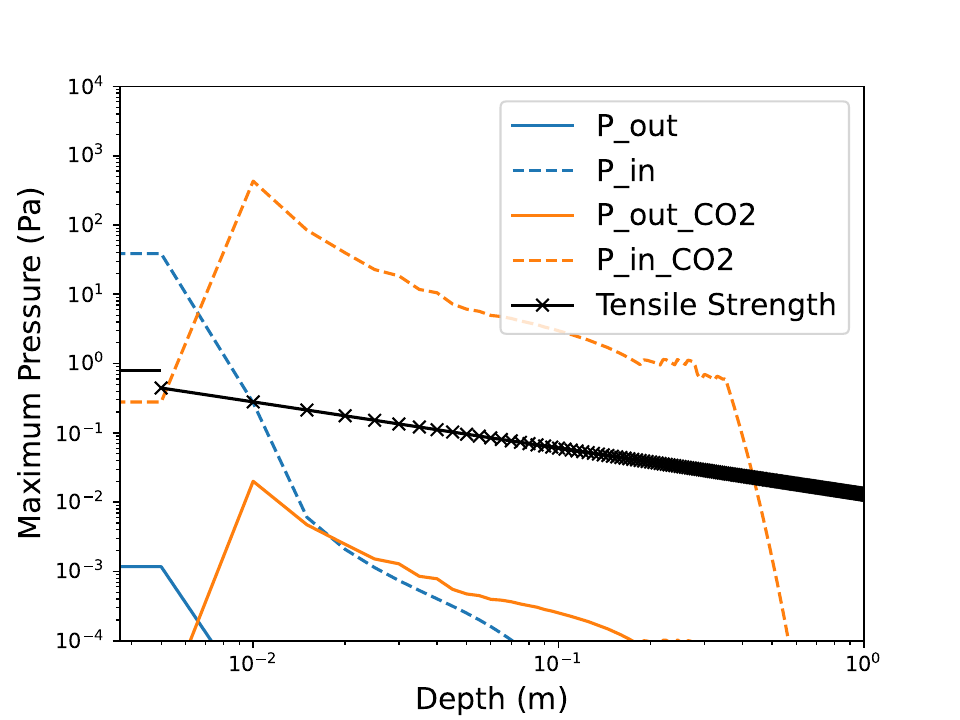}}%
\qquad
\caption{Simulation results for the \citet{Fulle2019} model non-WEB material with $\delta\approx\delta_{H_{2}O}=2$, $\delta_{CO_{2}}=100$. Top: average diurnal gas and dust emission rates, and their variation, during the Rosetta period as compared to the scaled observations. Bottom: Maximum outgassing pressure from different ice species in the subsurface, inside and in-between pebbles.}
\label{Plot:oldstyle}
\end{figure}

Figure \ref{Plot:oldstyle}, shows the results for $\delta\approx\delta_{H_{2}O}=2$, $\delta_{CO_{2}}=100$, i.e.~$1\%$ CO$_{2}$, relative to water ice. The water outgassing and dust ejection rates are very similar to the WEB case above: approximately 8 m of surface erosion into small dust-particles occurs between $\pm200$ days of perihelion and the peak water emission rate is still roughly matched. With an ice fraction of $1\%$, CO$_{2}$ drains quickly into the subsurface, reaching $z=30$ cm depth at aphelion and an outgassing rate somewhat lower than the observations away from perihelion. Near perihelion, when water activity is driving erosion, CO$_{2}$ spikes upwards whenever the removal of layers brings it closer to the surface, before falling quickly again as it drains away, resulting in an average depth of one or two cm and an average production-rate that can sometimes exceed that of water. The lower part of Fig.~\ref{Plot:oldstyle} shows that the CO$_{2}$ pressure between the pebbles never exceeds the tensile strength, meaning that there are no chunk ejections.

\subsection{Latitudinal variation}
\label{sec:results:latitude}

\begin{figure}
\subfloat{\includegraphics[scale=0.5]{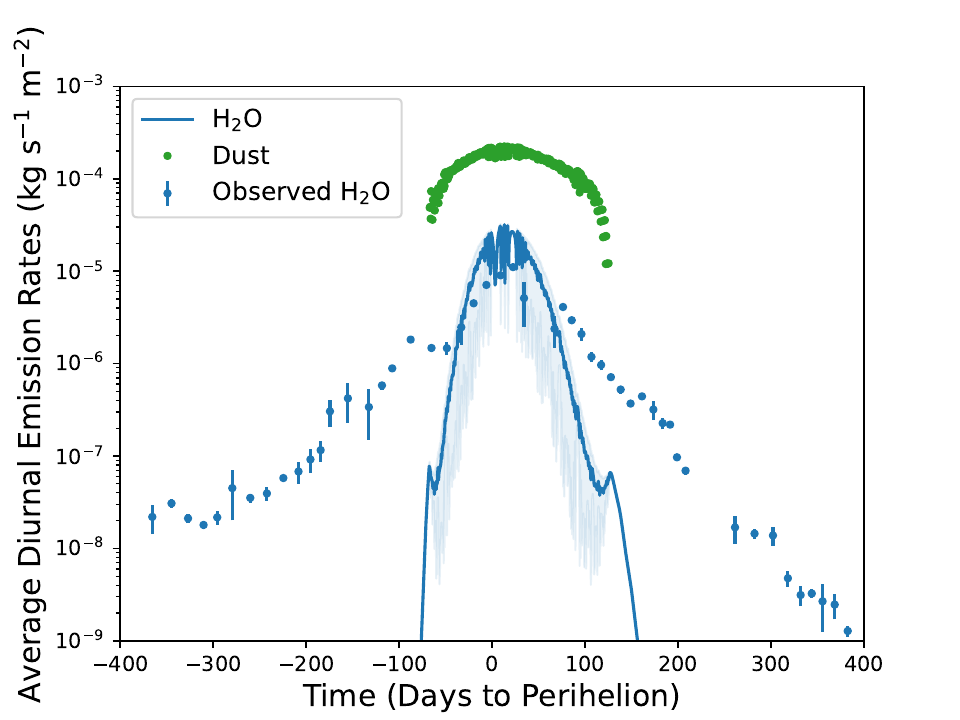}}%
\qquad
\subfloat{\includegraphics[scale=0.5]{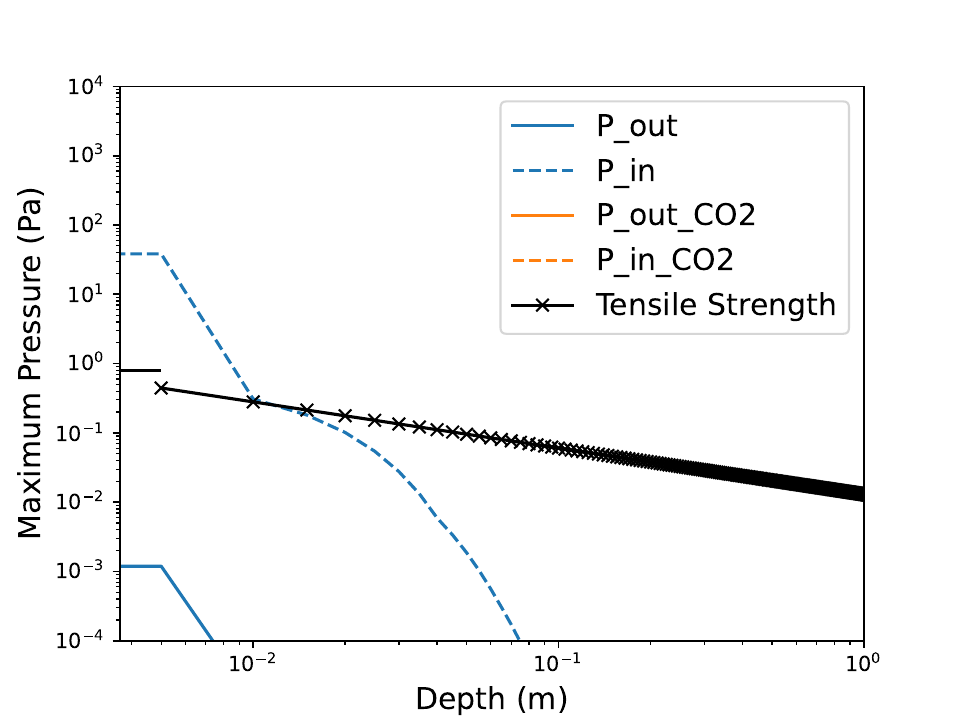}}%
\qquad
\caption{Simulation results for the \citet{Fulle2019} model WEB material, and a latitude of $-90^{\circ}$. Top: average diurnal gas and dust emission rates, and their variation, during the Rosetta period as compared to the scaled observations. Bottom: Maximum outgassing pressure from different ice species in the subsurface, inside and in-between pebbles.}
\label{Plot:WEBSouthPole}
\end{figure}

Insulation patterns on 67P are highly latitudinally dependent. In Figure \ref{Plot:WEBSouthPole}, we show the results for WEB material ($\delta=\delta_{H_{2}O}=2$) at the south pole, i.e.~latitude $-90^{\circ}$. Since we are only interesting in general trends, and do not wish to complicate the model by including local topography, we still only use the spherical shape model here.

The intense southern summer produces a higher peak production rate, with associated intense erosion of $\sim13$ m per orbit. As before, this comes entirely from sub-pebble erosion in the surface active layer. The peak observed water production rate at perihelion is now matched, but it can be seen that the outgassing of material at the south pole becomes negligible $\sim100$ d after perihelion as illumination fades into polar night. The lower part of the figure shows a similar maximum pressure with depth as for the equator. Very similar results to the equatorial case are also obtained for runs with $1\%$ CO$_{2}$ ice fraction with, as above, water dominating the activity and CO$_{2}$ pressure still not enough to exceed material strength.

\begin{figure}
\subfloat{\includegraphics[scale=0.5]{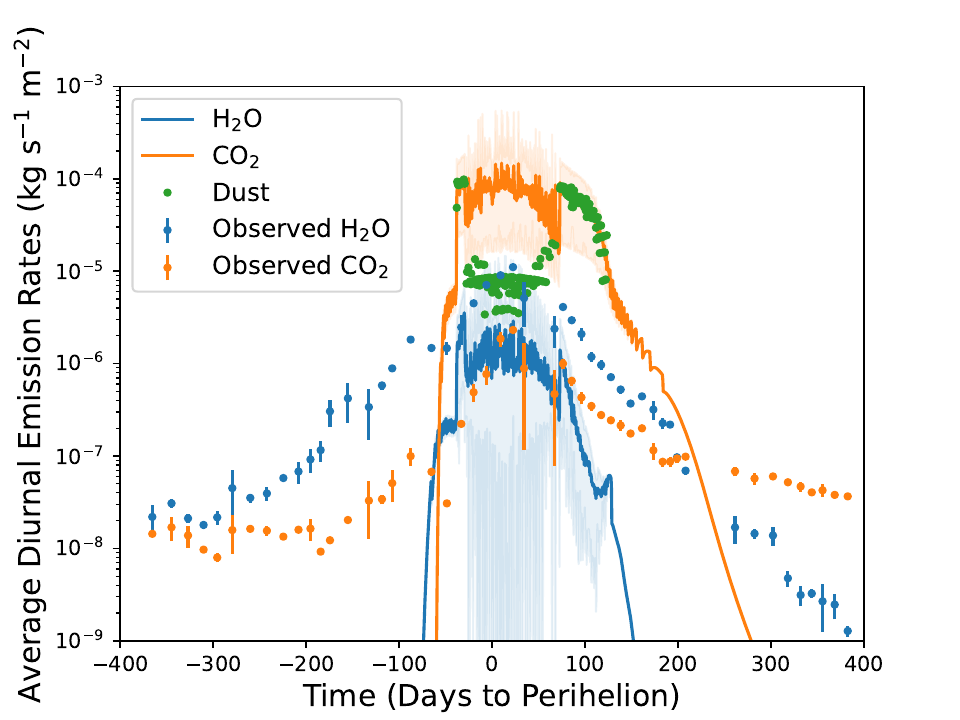}}%
\qquad
\subfloat{\includegraphics[scale=0.5]{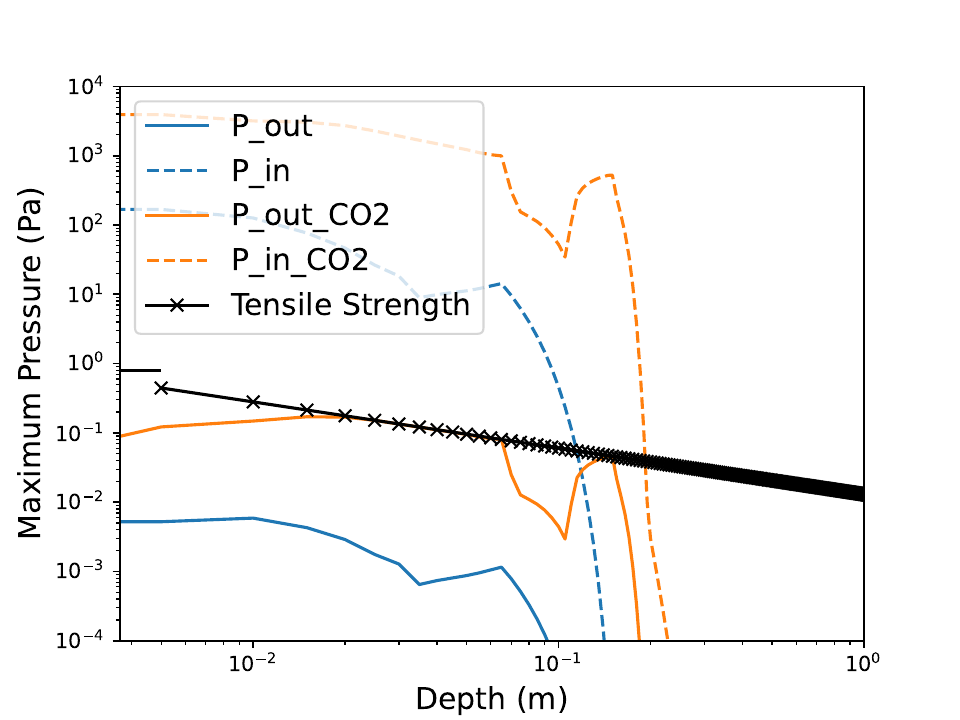}}%
\qquad
\caption{Simulation results for the \citet{Fulle2019} model non-WEB material with $\delta\approx\delta_{CO_{2}}=2$, $\delta_{H_{2}O}=50$, and a latitude of $-90^{\circ}$. Top: average diurnal gas and dust emission rates, and their variation, during the Rosetta period as compared to the scaled observations. Bottom: Maximum outgassing pressure from different ice species in the subsurface, inside and in-between pebbles.}
\label{Plot:SouthPole}
\end{figure}

Meanwhile, for  CO$_{2}$-rich non-WEB material the results are qualitatively different. For $\delta\approx\delta_{CO_{2}}=2$ at the south pole, Figure \ref{Plot:SouthPole} shows that the steep increase in insolation at the beginning of southern summer now means that pressure can quickly build up before CO$_{2}$ drains away. This allows it to exceed tensile strength in-between pebbles (see lower part of the plot), ejecting chunks up to 15 cm in size (roughly equal to the average observed by Rosetta; \citealp{Fulle2019b}), bringing fresh CO$_{2}$ and H$_{2}$O to the surface. CO$_{2}$, with it's higher volatility, then sublimates vigorously, ejecting small dust until it drains away from the first pebble and begins to decline. The top pebble then heats up until water begins sublimating, driving more dust erosion, but typically not enough to maintain the activity, leading it to also drain away. Temperatures then rise further and the heat propagating into the subsurface triggers another CO$_{2}$ ejection from a few layers deep ($\sim1-$a few cm), bringing both to the surface again, and the process repeats. This leads to an intermittent emission rate for both gas species, as average surface temperatures and ice contents jump up and down between those dominated by water and those by CO$_{2}$. The higher volatility of CO$_{2}$, however, means that, on average, its emission rate exceeds that of water. Dust is ejected as a mixture of sub-pebble and one-to-few pebble-sized particles.

As insolation declines, the dehydration rate of water decreases below the erosion rate, and a period of water driven sub-pebble erosion takes place, while CO$_{2}$ drains into the subsurface. Eventually, this too subsides, as H$_{2}$O also drains from the first pebble, halting further erosion. The relatively steep decrease in insolation compared to the equator, as the southern hemisphere retreats into its long winter night, then 'freezes in' the situation, with CO$_{2}$ still close to the surface (at a depth of $z=11$ cm), ready for the next orbit.

These results show some promise, in that WEB material can reproduce the observed perihelion water production-rate, and some chunk ejections by CO$_{2}$ are seen. However, the circumstances where these occur are limited, so we now return to the diffusion parameter to see if variations in $b$ can change the outcome.

\subsection{Low diffusivity case}
\label{sec:results:diffusivity}

As described in Section \ref{sec:method:diffusivity}, simulations and theory suggest that lower diffusivities may be appropriate for gas flowing between the pebbles. For pebble packing with porosities around the 0.4 expected for 67P, the curves in Figure \ref{Plot_Tempmodels_diffusivity} show $b\approx0.3 d_{p}$. We therefore reran the above equatorial cases with this value. The lower diffusivity allowed greater pressures to build up between pebbles, but in no cases did H$_{2}$O pressure exceed material strength, meaning that the results without CO$_{2}$ are almost identical to the  $b= d_{p}$ case above. Likewise, cases with CO$_{2}$, but with a quantity less than water, such as the $1\%$ CO$_{2}$ case from above, are almost identical to before (Fig.~\ref{Plot:oldstyle}), with an H$_{2}$O dominated outgassing, but insufficient CO$_{2}$ pressure to eject chunks. Interestingly, although the reduced diffusivity did change the pressure and outgassing rates slightly, the overall outgassing rate is very similar to before, due to the energy-balance naturally adjusting itself to the new conditions.

By contrast, in simulations of the non-WEB material with more CO$_{2}$ than water, the lower diffusivity allowed CO$_{2}$ pressures in-between the pebbles to overcome tensile strength, thus initiating chunk ejections.

\begin{figure}
\subfloat{\includegraphics[scale=0.5]{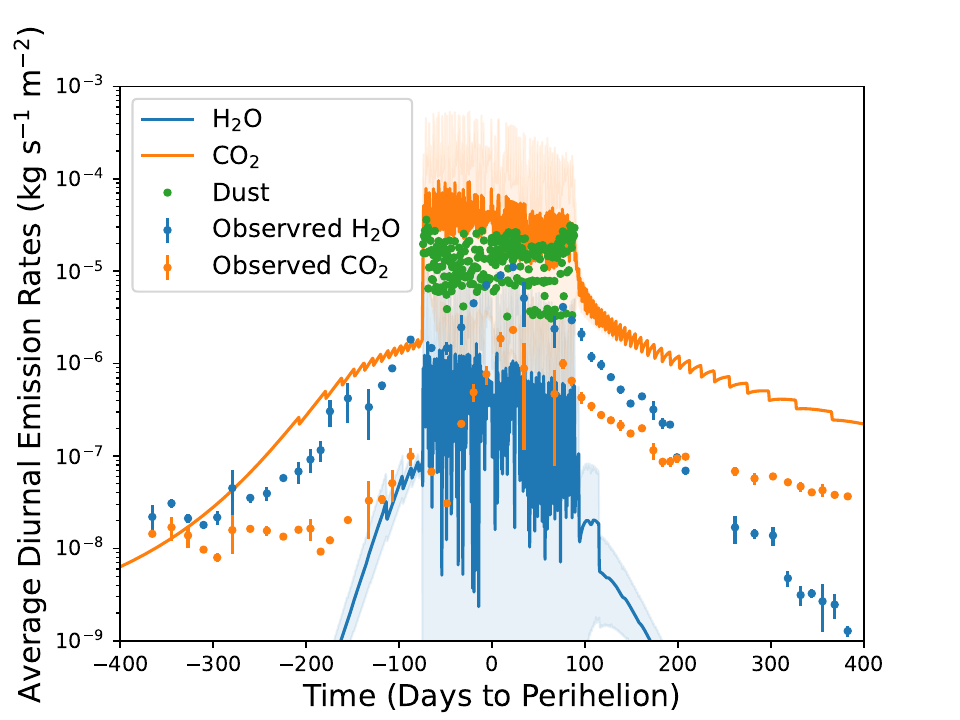}}%
\qquad
\subfloat{\includegraphics[scale=0.5]{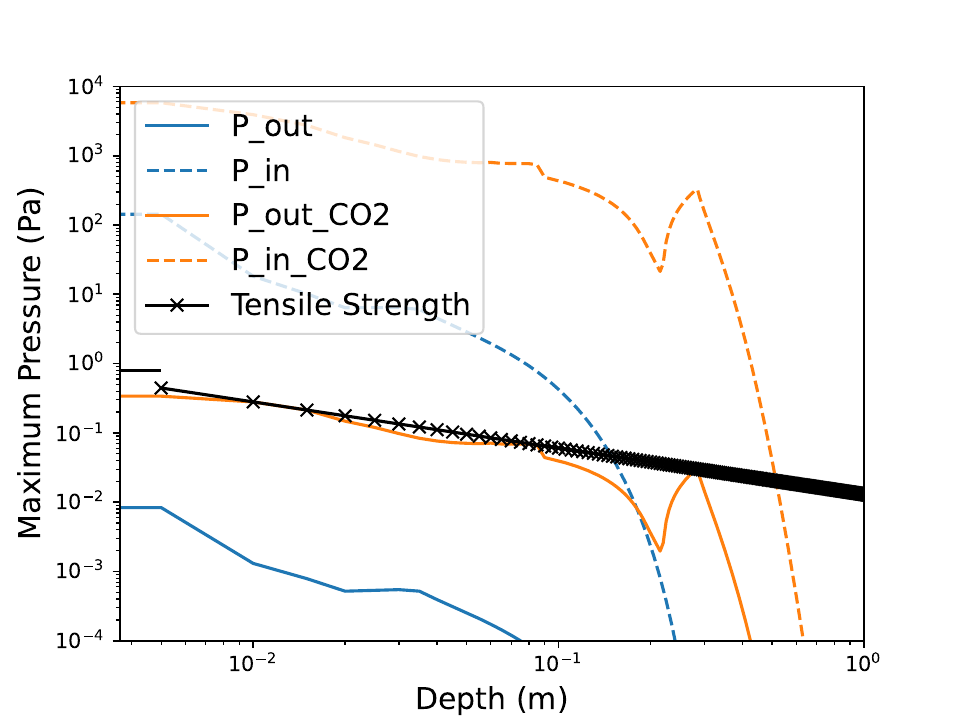}}%
\qquad
\caption{Simulation results for the \citet{Fulle2019} model non-WEB material with $\delta\approx\delta_{CO_{2}}=2$, $\delta_{H_{2}O}=50$, and $b=0.3d_{p}$. Top: average diurnal gas and dust emission rates, and their variation, during the Rosetta period as compared to the scaled observations. Bottom: Maximum outgassing pressure from different ice species in the subsurface, inside and in-between pebbles.}
\label{Plot:lowbnonWEB}
\end{figure}

The results for the $\delta\approx\delta_{CO_{2}}=2$ case are shown in Figure \ref{Plot:lowbnonWEB}. As in the higher diffusivity case, the small quantity of H$_{2}$O per layer swiftly drains away, retreating below the first pebble layer (to $z=1.5$ cm depth at aphelion) where the low diffusivity of the material above severely quenches its outgassing. Activity is seen within $\sim\pm100$ d of perihelion each orbit. Here, CO$_{2}$ pressure becomes great enough to eject a large, roughly 28 cm-sized chunk (see maximum pressure curve in the lower part of the plot), moving fresh CO$_{2}$ and H$_{2}$O directly to the surface. The situation is then the same as for the high-diffusivity case at the south pole (Fig.~\ref{Plot:SouthPole}): both volatiles sublimate in the top pebble or just below, driving sub-pebble dust erosion and ejections of one- or two-pebble layers to a total of around 4.5 m each orbit. The energy balance and outgassing are dominated by CO$_{2}$. Similar results were found for the $\delta\approx\delta_{CO_{2}}=10$ case from above; with differences in the maximum chunk size, total erosion depth, and exact outgassing rates, but still with CO$_{2}$ production exceeding H$_{2}$O. Thus, while the lower diffusivity helps with generating enough pressure to launch large chunks, we run into the same problem of 'run-away' CO$_{2}$-driven activity also seen for the higher diffusivity case at the south pole.

Finally, for completeness, we also investigated the original model of \citet{bischoff2023} (i.e.~not using the internal pebble pressure described in Section \ref{sec:method:pinside}), with the lower diffusivity $b=0.3 d_{p}$. The lower diffusivity means that higher pressures can build up, but for water this was still insufficient to overcome the tensile strength at the equator. For very high CO$_{2}$ concentrations (such as the $\delta\approx\delta_{CO_{2}}=2$ non-WEB material), CO$_{2}$ pressure did exceed tensile strength, leading to around 4.2 m of pebble- and above-sized chunk ejections per orbit, up to maximum chunk size of $\sim38$ cm. However, as above, this led to CO$_{2}$ at, or near the surface during the entire active period, resulting in very high CO$_{2}$ emission rates and relatively low H$_{2}$O outgassing.

\subsection{Numerical resolution}
\label{sec:results:resolution}

Before discussing the implications of this modelling, we must first confirm that the results are not influenced by the numerical resolution of the time and depth steps used in the discretisation. As described in Section \ref{sec:method:strength}, we do not attempt to simulate the detailed inner workings of a pebble here, so our minimum depth step is fixed at half a pebble radius. Changing the numerical resolution therefore also changes the pebble size, which has an influence on the pebble outgassing rate via Eqn.~\ref{eq:q}. We tried several tests with different sized pebbles and found very similar results to those shown above as the total outgassing and pressure again adjusted themselves to the new energy balance. The exception was for small pebble sizes ($\sim1$ mm): here the per-pebble outgassing rate was large enough for the surface pebble to drain quickly, quenching activity. The results of reducing pebble size therefore tend towards those of the non-pebble models \citep{bischoff2023}. In terms of the time resolution, our default time-step (10 s) is ten times smaller than the minimum used by \citet{Gundlach.2020}. We also tested a run  with a time-step of 1 s for one orbital period, for the equatorial non-WEB case ($\delta_{CO_{2}}=2$, $\delta_{H_{2}O}=50$), and found near identical results to the default case.

\section{Discussion}
\label{sec:discussion}

From the above results we can see that ejection of dust particles from within pebbles is readily achieved in the \citet{Fulle2019} model, but that larger chunks are much more difficult, and only ejected when there is a lot of CO$_{2}$ (greater than or equal to the water mass), and under specific conditions (low diffusivities or intense insolation at southern summer). In addition, no chunk ejections are seen when there is ongoing water-driven erosion of the first pebble. This is because H$_{2}$O sublimation consumes most of the incoming energy, preventing the heat-wave from significantly warming the interior. These results argue against the elaboration of the basic \citet{Fulle2019} model described in \citet{Fulle2021}, wherein non-WEB material keeps eroding and emitting H$_{2}$O gas at the same rate as WEB material during the whole period that it dehydrates, before smoothly being replenished by continuous CO$_{2}$-driven chunk ejection. Instead, we see non-WEB material experiencing a decreasing H$_{2}$O emission with time as water drains. This can be followed some time later by a CO$_{2}$ driven chunk ejection, with the delay caused by the time needed for the thermal wave to propagate to the CO$_{2}$ depth, beginning only once H$_{2}$O sublimation drops off and surface erosion ceases. After the chunk ejection, CO$_{2}$ ice is then directly present within the surface pebble for some time before it drains, leading to extremely large CO$_{2}$ outgassing rates. Average CO$_{2}$ outgassing then exceeds that of H$_{2}$O for CO$_{2}$-rich material. 

From this, it seems unlikely that material with more CO$_{2}$ than H$_{2}$O ice-content dominates the outgassing surface area of comet 67P. If this were the case, as in the $\sim90\%$ surface fraction of CO$_{2}$-rich non-WEB material suggested in \citet{Ciarniello2022}, then our numerical modelling would suggest a much larger CO$_{2}$ production than that measured by Rosetta. We note that the problem of 'run-away' CO$_{2}$-driven activity is not limited to the model of \citet{Ciarniello2022}, and that any time CO$_{2}$ drives the ejection of dust (whether in chunks, or sub-pebble particles) it is then automatically near the surface and experiencing large sublimation rates. This is also seen in the CO$_{2}$-driven models in the first paper \citep{bischoff2023}.

\citet{Gundlach.2020}, by contrast, did manage to achieve a good ratio of CO$_{2}$ to H$_{2}$O outgassing simultaneously with chunk ejection, but only for models with less CO$_{2}$ than water, and under the specific conditions of constant illumination at perihelion. \citet{bischoff2023} struggled to extend this result to time-varying conditions with a day and night cycle and seasonally differing illumination. Meanwhile, the pebbles simulated in both these works did not contain the volatiles inside of them, thus differing from the description of \citet{Fulle2020} used here. Nevertheless, the positive results of \citet{Gundlach.2020} suggest that CO$_{2}$ might be located outside of the pebbles, in order for it to trigger chunk ejections from depth, while the H$_{2}$O may still be contained inside the pebbles, in order for it to erode them into fine dust. We investigate this scenario in the globally summed models below, while noting that a process to transfer CO$_{2}$ from inside pebbles in dynamically new comets \citep{Fulle2022} to outside of them in 67P is currently unknown. 

\subsection{Global models}

\begin{figure}
\subfloat{\includegraphics[scale=0.5]{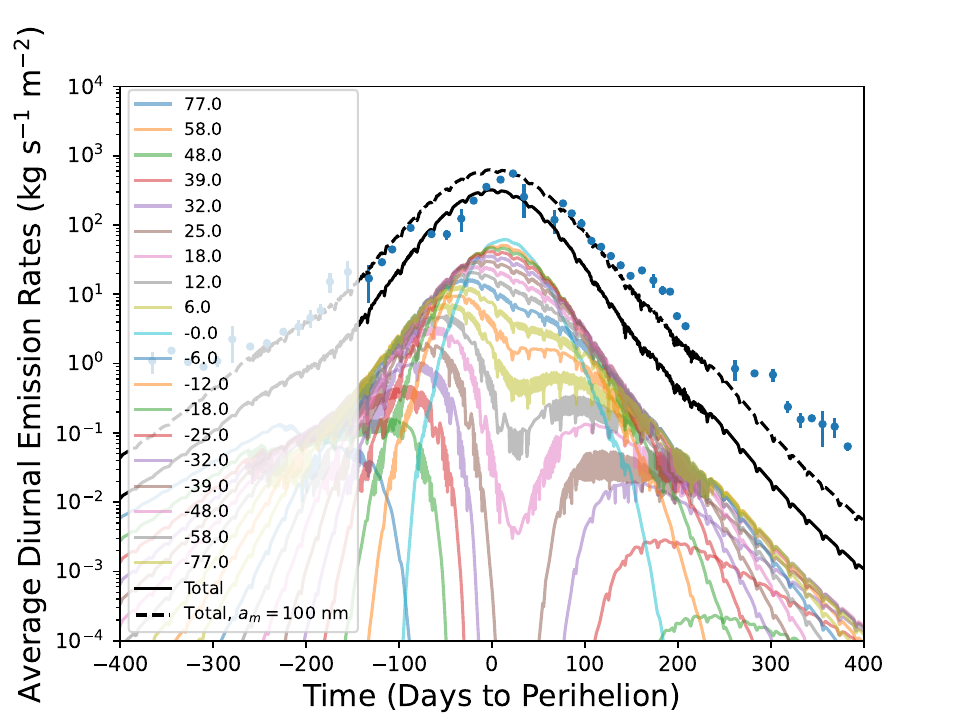}}%
\qquad
\subfloat{\includegraphics[scale=0.5]{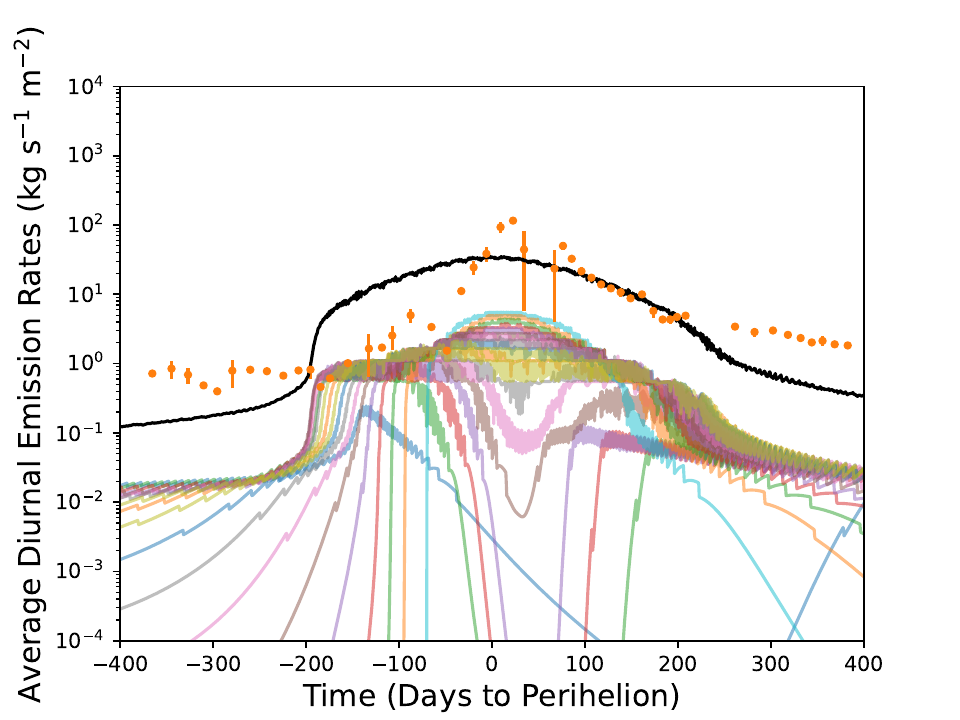}}%
\qquad
\subfloat{\includegraphics[scale=0.5]{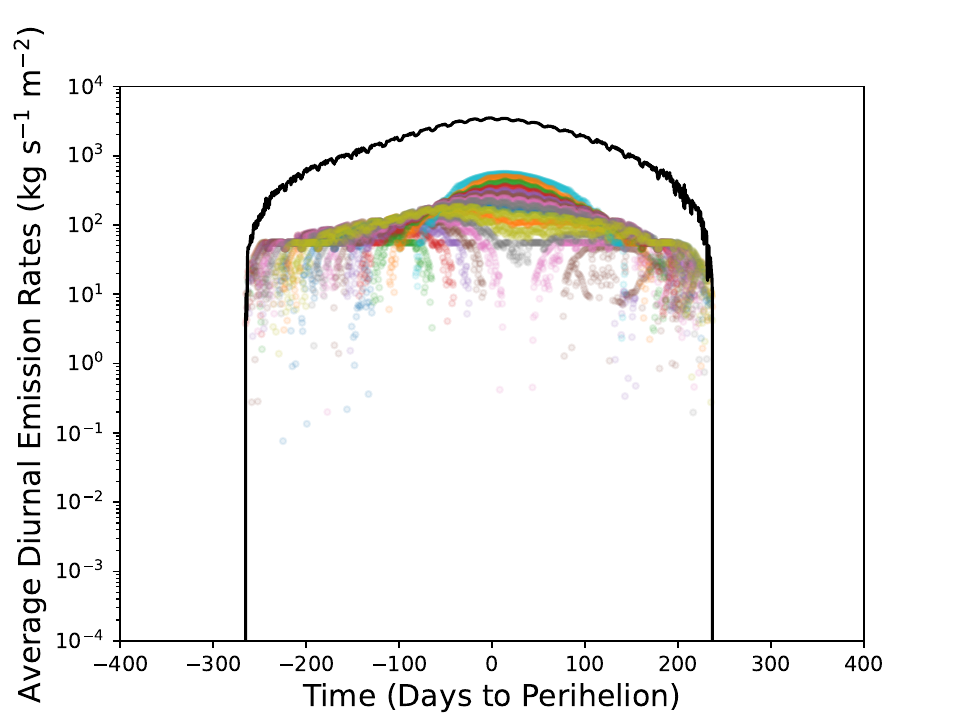}}%
\caption{Average diurnal emission rates over the full comet surface for the \citet{Fulle2019} model with $\delta\approx\delta_{H_{2}O}=2$, $\delta_{CO_{2}}=100$. Top: H$_{2}$O, middle: CO$_{2}$, bottom: dust. Each of the 19 simulated latitudes is shown in a different colour, which are consistent across the plots.}
\label{Plot:summed}
\end{figure}

In order to constrain the areal fractions of the various material types, we made use of the parallel computing aspect of the code to run simulations over a number of latitudes on the spherical comet before summing their area-weighted contributions. We present two such models below. First, figure \ref{Plot:summed} shows the results for a full WEB surface ($\delta\approx\delta_{H_{2}O}\approx2$, with the addition of $1\%$ CO$_{2}$). We note that a surface completely covered in WEBs is not realistic given the constrains from the changing nucleus colour \citep{Ciarniello2022} and bright-spot concentrations \citep{Fornasier2023}, but wish to investigate its outgassing and dust emitting potential.

As can be seen in the top panel, peak water production at perihelion is well matched by a combination of equatorial and southern hemisphere outgassing. Before and after perihelion, however, the contribution of the south declines rapidly, and the northern hemisphere and equatorial regions together struggle to supply enough water outgassing. This is due to the effects of thermal inertia combined with the high damping (i.e.~low per-pebble emission rate) inherent in the \citet{Fulle2019} model. The gas production away from perihelion is lower here than that shown in \citet{Ciarniello2023} because heating up the material in the time-dependent solution reduces the energy available for sublimation as compared to the instantaneous response of the latter. This thermal inertia problem is inescapable in a full numerical solution, but this is not to say that the model is incapable of reproducing the observations. In Figure \ref{Plot:summed} we also show the summed total water-production for an identical run with the minimum pore-size within the pebbles set to $a_{m}=100$ nm, rather than 20 nm. This has the effect of increasing diffusivity inside the pebbles, and hence their production rate, and could be justified by the variation in diffusivity shown in Figure \ref{Plot_Tempmodels_diffusivity}. With a higher per-pebble emission rate, the peak water production near perihelion is now slightly exceeded, but the curve fits the increasing and decreasing outgassing between $\sim\pm200$ days much better. CO$_{2}$ and dust production are essentially unaltered. Before 300 days prior to perihelion, the observed rate is still not met, which may be due to a contribution from extended sources, as posited by \citet{Ciarniello2023}, while from 200 days after perihelion there remains a discrepancy. This could also be due to extended sources, or a reactivation of volatile-rich fallback material transported to the north from the active south. We do not seek to explain these scenarios, including any pore-size variation, in detail here, merely to show that the modelled \citet{Fulle2019}-style water-production curve can be roughly consistent with the water-production data.

When considering the CO$_{2}$ production (middle panel), we can see that this case, with a small quantity of CO$_{2}$, can roughly reproduce the peak observed rate. Reproducing the whole time-dependent curve is again more challenging, however, with the shape of our modelled curve not matching the observations. It is likely that latitudinal variations in CO$_{2}$ content, i.e.~spatial variation, could improve the situation, but we leave this modelling for future work. Overall CO$_{2}$ production here is somewhat low, i.e.~integrating the curves with time gives a total CO$_{2}$ mass-loss of $6.7\times10^{8}$ kg, compared to the $1.0 \pm 0.2\times 10^{9}$ estimated by \citet{Laeuter2020} for the Rosetta period. For water, the two production curves in the top part of the plot result in 2.8 or $6.2\times10^{9}$ kg per orbit, bracketing the measured $4.5\pm0.6\times10^{9}$ kg. 

Conversely, total dust-production, as shown in the bottom curve, is overestimated in the \citet{Fulle2019} model. This is a known feature of the model, and \citet{Fulle2021, Ciarniello2022}, etc.~invoke an active fractional area of eroding WEB material on the order of a few percent in order to reduce the flux by this fraction (which also matches the WEB fraction implied by the nucleus colour and bright-spot constraints). Our total modelled dust mass-loss is $6.8\times10^{11}$ kg. By comparison, the entire observed total mass-loss, including volatile outgassing and solids, is measured by \citet{Patzold} as $10.5 \pm 3.4\times10^{9}$ kg, resulting in a dust mass-loss of $\approx5\times10^{9}$ kg, when subtracting the volatile totals mentioned above \citep{Laeuter2020}. Thus, the curve in Fig.~\ref{Plot:summed}, overestimates dust production by around 130 times, requiring a WEB active fraction of around $0.7\%$. Some portion of the ejected dust will return to the surface as fallback, which would help reduce the modelled curve relative to the measurements and allow the active fraction to be higher. We do not estimate fallback fractions in our model (which vary in time and space, depending on particle sizes and outgassing rates, and can be a significant fraction: see \citealp{Marschall2020}), but do note that all the particles lifted here are smaller than the pebble size of 1 cm, and so a significant fraction should escape. For steep power-law size distributions with exponents greater than 3, \citet{Marschall2020} estimate fallback fractions $<50\%$ and down to a few percent, depending on gas production, so that our WEB active fraction cannot be more than around $1.5\%$.

Based on the argument regarding the results of \citet{Gundlach.2020}, above, we also consider a model where the CO$_{2}$ is located outside of the pebbles. This hybrid model therefore uses equations \ref{eq:Pin}, \ref{eq:q}, \ref{eq:Pout}, and \ref{eq:qout} for calculating H$_{2}$O pressure (inside and outside the pebbles) and emission rate, and the standard equations of \citet{Gundlach.2020} and \citet{bischoff2023} for CO$_{2}$. Zero CO$_{2}$ pressure inside the pebbles is assumed. Initial experiments with small quantities of CO$_{2}$ (the 1$\%$ case above) resulted in little difference in overall outgassing rates with this hybrid model, as the depth of the sublimation front and emission rate again adjusted to the new energy balance without CO$_{2}$ pressure exceeding the material strength.

\begin{figure}
\subfloat{\includegraphics[scale=0.5]{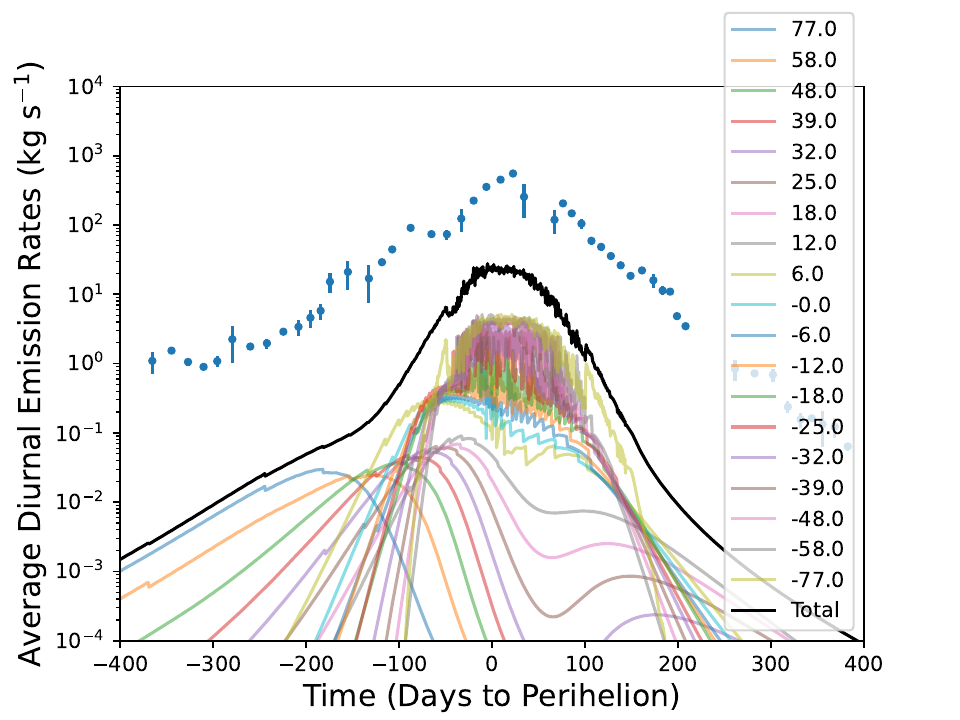}}%
\qquad
\subfloat{\includegraphics[scale=0.5]{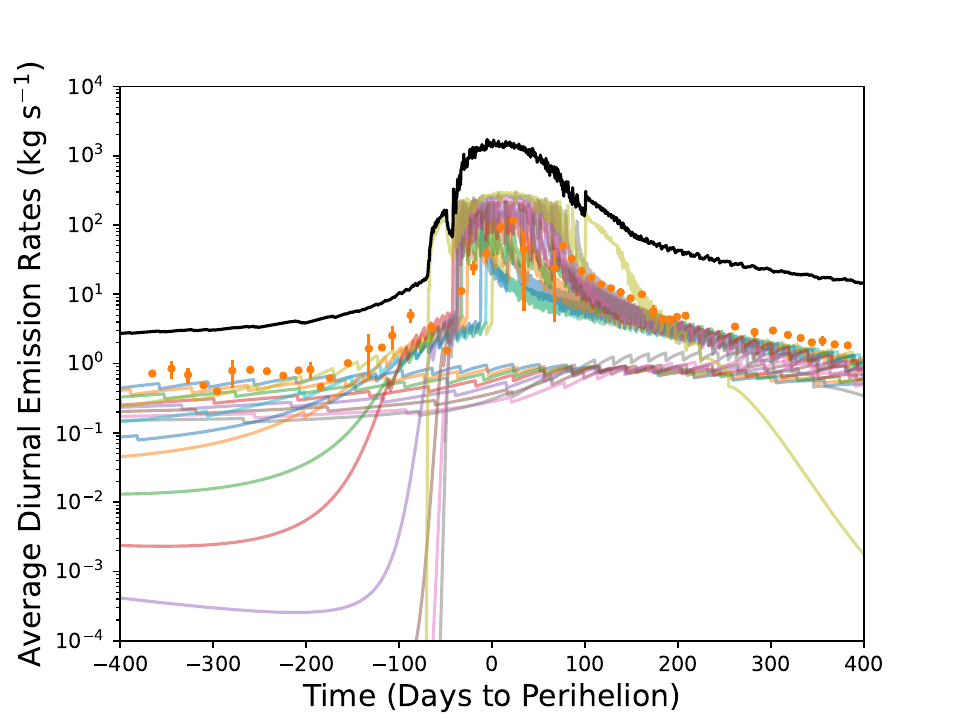}}%
\qquad
\subfloat{\includegraphics[scale=0.5]{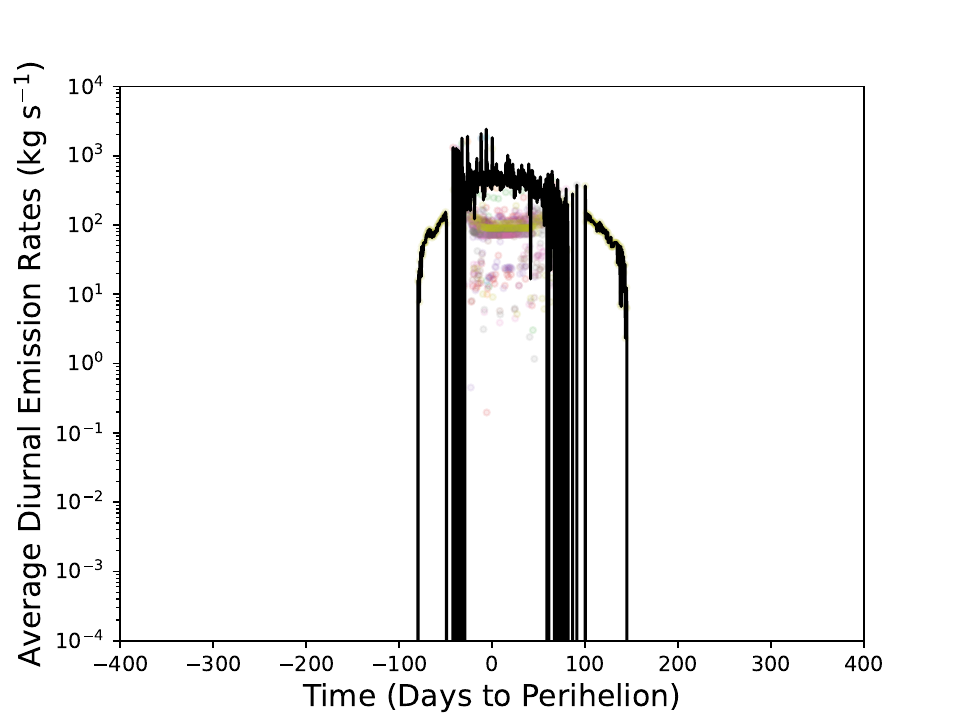}}%
\caption{Average diurnal emission rates over the full comet surface for a hybrid model with $\delta\approx\delta_{CO_{2}}=2$, $\delta_{H_{2}O}=50$. Top: H$_{2}$O, middle: CO$_{2}$, bottom: dust. Each of the 19 simulated latitudes is shown in a different colour, which are consistent across the plots.}
\label{Plot:summed_hybrid}
\end{figure}

For large quantities of CO$_{2}$, however, the results are slightly different. Figure \ref{Plot:summed_hybrid} shows the global emission rates for the ice-rich non-WEB case,  $\delta\approx\delta_{CO_{2}}=2$, $\delta_{H_{2}O}=50$. By not being inside the pebbles, CO$_{2}$ emission is less damped than before, and consequently it drains deeper into the subsurface and can reach large pressures deep down ($\sim10-30$ cm) to trigger chunk ejections. H$_{2}$O behaves as before, eroding the surface pebble into fine dust whenever it is present there, before draining away due to its small concentration. The result is quite similar to the previous experiments, with a water outgassing rate roughly an order of magnitude below the observations (top panel) and intermittent ejection of large chunks and small dust (bottom panel). The CO$_{2}$ emission (middle panel) is again too large, due to its very high concentration, however the shape of the curve now resembles the observations. This is particularly true from $\sim100$ days after perihelion, when dust ejections have stopped and the CO$_{2}$ is draining into the subsurface. Scaling the CO$_{2}$ curve down to around $10\%$ active surface would reproduce the observations quite well, though of course this would leave the H$_{2}$O much too low by roughly two orders of magnitude.

\subsection{Localised ejection of chunks}

Overall then, it is difficult to constrain the WEB and/or CO$_{2}$-rich area fractions from this analysis, except to say both must be significantly less than the total surface area of 67P. For the former, this is to prevent the flux of small, sub-pebble dust becoming too large (as well as to satisfy the surface colour and bright-spot constraints), while in the latter, the constraint is posed by the measured CO$_{2}$ outgassing. If both fractions are small, then inert, low volatile-content material would cover the majority of the surface. This would be at odds with most Rosetta measurements, however, that suggest water and dust ejection from everywhere. It would also further reduce the total water-outgassing curve which, as Figure \ref{Plot:summed} shows, would mean that the \citet{Fulle2019} model struggled to reproduce the observations without an increased diffusivity inside the pebbles. A solution could be if the proportion of fallback was much higher than estimated above. If, for example, the ejection speed of small particles from within eroding pebbles is very low (perhaps because they suffer immediate collisions with their neighbours, robbing them of momentum), then they would not escape 67P's gravity, and would fall back locally. This would reduce the modelled dust-ejection and erosion rates, and could bring them back into comparison with the data. The WEB fraction, and total area eroding by water-driven pebble erosion, would then be high in this model, which would bring it into conflict with the constraints from nucleus colour and bright-spot concentrations.

CO$_{2}$ would then mostly be outgassed from small localised areas of high concentration, with the potential addition of a low background emission coming from below water-eroding areas predominantly in the northern hemisphere, similar to the $1\%$ CO$_{2}$-case above (this may account for the 'base' CO$_{2}$ emission rate described in \citealp{Ciarniello2023}). The relative balance of these two is difficult to compute without further, spatially resolved modelling, but suffice it to say that the surface area eroding by CO$_{2}$ activity should be smaller than that eroding by water at any one time, so as to minimise the outgassing of the former as compared to the latter. Water outgassing, together with dust emission, would then be separated in time and space from large chunk emission, which we would expect to see occurring in clusters from specific locations, and associated with enhanced CO$_{2}$ outgassing. These would typically follow the pattern of one large chunk, followed by small dust and outgassing of both ice species which declines with time, followed by a few smaller chunks. Water emission (together with dust erosion) could be interspersed with this at the same location over seasonal timescales if there are changes in composition with depth, provided there is sufficient erosion by each species to expose the other. Time and spatially varying activity patters then become very complex, and dependent on the exact depth profile. Overall though, the high water-driven erosion rate experienced in the southern hemisphere should expose more fresh CO$_{2}$-rich material per unit time, leading to the observed increase in CO$_{2}$ emissions at perihelion; while erosion in the north only exposes more CO$_{2}$-depleted fallback material, limiting the outgassing of this species and the ejection of chunks here.

The separation of dust and chunk emissions described above may be very localised and difficult to resolve with the rather broad-scale Rosetta measurements. For example, \citet{Laueter2019} estimate surface emission of  H$_{2}$O and CO$_{2}$ from ROSINA gas density measurements, and find concentrations of both in a set of large, km-scale areas. This could support the idea of the two species being collocated, but might also represent separate concentrations of both at the metre-scale. Nonetheless localised chunk ejection is tentatively supported by the detection of individual boulder and chunk ejections from particular locations by the OSIRIS camera \citep{Lemos2023, Lemos2024, Pfeifer2024, Shi2024}. A more complete survey of 67P's variation in seasonal \citep{Marschall2020} and diurnal \citep{Gerig2020} emission of dust and chunks of various sizes would help to constrain these models.

Finally, we note that modelling of 67P's outgassing driven non-gravitational accelerations and torques points to a variation in the insolation response of different morphological terrain types and across differently orientated parts of the nucleus \citep{Attree2024}. This does not support uniform material properties everywhere, suggesting a variation in local outgassing behaviour, again making it difficult to constrain materials properties with single 1D thermophysical model. It must also be remembered that our modelling here assumes locally flat surfaces on a spherical nucleus and therefore ignores topography. Inclined surfaces will receive differing amounts of radiative energy input (which could be more or less than here, due to shadowing and self-heating by other surfaces), which may change the outgassing response and the balance between ejections and volatile draining.

\section{Conclusions}
\label{sec:conclusions}

We have extended the thermophysical model of \citet{bischoff2023} to include the description by \citet{Fulle2019} of substructure and pressure buildup within the pebbles making up cometary nuclei. We discuss its numerical implementation (Section \ref{sec:TPM:Basis}) and the questions of cometary material strength and gas diffusivity specifically in Sections \ref{sec:method:strength} and \ref{sec:method:diffusivity}. We then test various dust-to-ice ratios of H$_{2}$O and CO$_{2}$ ice, in order to simulate the material inside and outside of the water enriched bodies (WEBs) proposed by \citet{Ciarniello2022} (Section \ref{sec:results}).

Overall, we find that the model of \citet{Fulle2019} can reproduce the peak water flux observed by Rosetta at 67P at perihelion, whilst also generating continuous dust-lifting activity that replenishes the surface over at-least three orbital cycles. The total modelled erosion and ejected dust-mass are larger than the observations, however, necessitating an active area smaller than the total comet surface area, or very large quantities of dust fallback. Further, the addition of time-resolved heat-flow, versus the static equilibrium model of \citet{Fulle2019} and \citet{Fulle2020}, leads to a reduced water production-rate away from perihelion, which must be compensated by additional factors such as extended water sources or a different diffusivity within the pebbles to originally assumed.

When simulating the CO$_{2}$-rich non-WEB material described in \citet{Fulle2021} and \citet{Ciarniello2022}, we find that we can only eject large chunks under specific conditions (e.g.~low diffusivities of $b=0.3d_{p}$ between the pebbles, or intense insolation at southern summer), whilst we also find a CO$_{2}$ outgassing rate that greatly exceeds the Rosetta measurements. We find this to be a problem with all thermophysical models where CO$_{2}$ drives erosion: in these cases, its outgassing always exceeds that of water.

We therefore come to the conclusion that CO$_{2}$ driven erosion cannot dominate 67P's surface area, otherwise Rosetta would have observed much higher CO$_{2}$ production-rates. Considering the heat-wave propagation into subsurface, it is also very difficult to eject chunks from deep whilst simultaneously eroding and outgassing from the surface layer. We suggest that ejection of chunks by CO$_{2}$ must, therefore, be a localised process, occurring in particular spots on the cometary surface, separated from surface erosion and water emission in space or time (i.e.~by a horizontal or vertical separation of CO$_{2}$-rich material from H$_{2}$O-rich material in the subsurface). The area fraction of each material type exposed on the surface will then vary with time, as suggested by \citet{Ciarniello2022}. Constraining the proportion of CO$_{2}$-rich versus water-rich material, and matching the overall Rosetta outgassing measurements, is then extremely difficult with 1D thermal models of single patches on a spherical nucleus, even at multiple latitudes as we applied in Section \ref{sec:discussion}.

Overall then, important questions about the cometary activity mechanism still remain unanswered, and it is still a struggle to explain how dust of various sizes is broken off and lifted from the surface. Moving from the local simulation of individual patches to the global production rates measured by Rosetta also remains challenging. Nonetheless, numerical modelling such as the above places important constraints on the properties of the WEB and non-WEB material if they are present, while also highlighting the importance of the grain assembly and structure (particularly in determining the porosity and diffusivity at various scales) to the cometary activity mechanism. It is hoped that further analysis of the spatial and temporal pattern of dust and chunk ejections from Rosetta data, combined with additional modelling and laboratory experiments, may improve our understanding. 

\section*{Acknowledgements}

We thank the reviewer for their thorough and detailed review which helped improve the manuscript. N.A.’s contributions were made in the framework of a project funded by the European Union’s Horizon 2020 research and innovation programme under grant agreement No 757390 CAstRA, and acknowledges financial support from project PID2021-126365NB-C21 (MCI/AEI/FEDER, UE) and from the Severo Ochoa grant CEX2021-001131-S funded by MCI/AEI/10.13039/501100011033. D.B. and J.B. thank DFG for funding project BL 298/27-1. 
This research was supported by the International Space Science Institute (ISSI) in Bern, through ISSI International Team project \#547 (Understanding the Activity of Comets Through 67P's Dynamics).

\section*{Data Availability}
The data underlying this article will be shared on reasonable request to the corresponding author.




\bibliographystyle{mnras}
\bibliography{paper_references} 


\bsp	
\label{lastpage}
\end{document}